# 50 questions on Active Assisted Living technologies

## Global edition

cost — EUROPEAN COOPERATION IN SCIENCE & TECHNOLOGY

Funded by the European Union

goodbrother

## Suggested citation

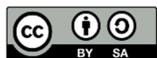

This publication is based upon work from COST Action GoodBrother – Network on Privacy-Aware Audio- and Video-Based Applications for Active and Assisted Living (CA19121), supported by COST (European Cooperation in Science and Technology).

COST (European Cooperation in Science and Technology) is a funding agency for research and innovation networks. Our Actions help connect research initiatives across Europe and enable scientists to grow their ideas by sharing them with their peers. This boosts their research, career and innovation.

https://www.cost.eu/

https://www.cost.eu/actions/CA19121

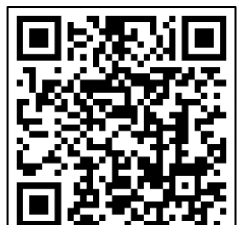

Get the electronic version of this document

**Table of contents**

## Welcome

Welcome to this booklet on Active Assisted Living (AAL) technologies, created as part of the GoodBrother COST Action, which has run from 2020 to 2024. COST Actions are European research programs that promote collaboration across borders, uniting researchers, professionals, and institutions to address key societal challenges. GoodBrother focused on ethical and privacy concerns surrounding video and audio monitoring in care settings. The aim was to ensure that while AAL technologies help older adults and vulnerable individuals, their privacy and data protection rights remain a top priority.

This booklet is designed to guide you through the role that AAL technologies play in improving the quality of life for older adults, caregivers, and people with disabilities. AAL technologies offer tools for those facing cognitive or physical challenges. They can enhance independence, assist with daily routines, and promote a safer living environment. However, the rise of these technologies also brings important questions about data protection and user autonomy.

Through practical case studies, this booklet explores real-world scenarios where AAL technologies are applied. They offer insights into their benefits and challenges. It discusses situations where older adults have regained their sense of independence. For instance, by using fall detection systems or health monitoring devices. These stories offer a human perspective on the technology, showing how it can enhance well-being while raising questions about privacy and ethical use.

The booklet also highlights the importance of informed decision-making when choosing and implementing AAL systems. It emphasises the need for these technologies to balance safety and convenience with respect for individual rights. You will find information on key considerations. These include how to select appropriate devices, manage privacy settings, and ensure the technology fits seamlessly into daily life.

The booklet, in addition to these case studies, addresses important topics. These include data security, ethical considerations, and the role of European regulations like the General Data Protection Regulation (GDPR) in protecting user privacy. Understanding these issues is essential for ensuring that AAL systems are not only effective but also respectful of users' rights and dignity.

This resource is intended for a wide audience, including end users, caregivers, healthcare professionals, and policymakers. It provides practical guidance on integrating AAL technologies into care settings while safeguarding privacy and ensuring ethical use. The insights offered here aim to empower users and caregivers to make informed choices that enhance both the quality of care and respect for personal autonomy.

Francisco Florez-Revuelta

Chair of the GoodBrother COST Action

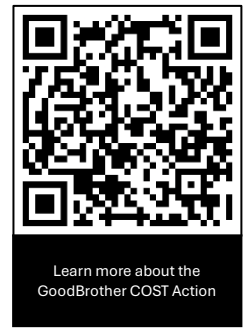

Learn more about the GoodBrother COST Action

**Contributors (in alphabetical order)**

Alin Ake-Kob, Nord University, Norway

Pau Climent-Perez, Universidad de Alicante, Spain

Paulo Coelho, Instituto Politecnico de Leiria, Portugal

Liane Colonna, Stockholm University, Sweden

Laila Dahabiyeh, The University of Jordan, Jordan

Carina Dantas, SHINE 2Europe, Lda, Portugal

Esra Dogru-Huzmeli, Queen's University at Kingston, Canada

Hazım Kemal Ekenel, Istanbul Technical University, Türkiye

Francisco Florez-Revuelta, Universidad de Alicante, Spain

Aleksandar Jevremovic, Singidunum University, Serbia

Nina Hosseini-Kivanani, Université du Luxembourg, Luxembourg

Aysegul Ilgaz, Akdeniz University, Türkiye

Mladjan Jovanovic, Singidunum University, Serbia

Andrzej Klimczuk, SGH Warsaw School of Economics, Poland

Maksymilian M. Kuźmicz, Stockholm University, Sweden

Petre Lameski, Ss. Cyril and Methodius University, North Macedonia

Ferlanda Luna, Universidade de Coimbra, Portugal

Natália Machado, SHINE 2Europe, Lda, Portugal

Tamara Mujirishvili, Universidad de Alicante, Spain

Zada Pajalic, University of South-Eastern Norway, Norway

Galidiya Petrova, Technical University of Sofia, Bulgaria

Nathalie G.S. Puaschitz, VID Specialized University, Norway

Maria Jose Santofimia, Universidad de Castilla-La Mancha, Spain

Agusti Solanas, Universitat Rovira i Virgili, Spain

Wilhelmina van Staalduinen, AFEdemy, Academy on age-friendly environments in Europe BV, Netherlands

Ziya Ata Yazıcı, Istanbul Technical University, Türkiye

**Abbreviations**

| | |
|---|---|
| **AAL** | Active Assisted Living |
| **ADLs** | Activities of Daily Living |
| **AI** | Artificial Intelligence |
| **CO2** | Carbon Dioxide |
| **EU** | European Union |
| **FAQ** | Frequently Asked Questions |
| **GDPR** | General Data Protection Regulation |
| **GPS** | Global Positioning System |
| **ICT** | Information and Communications Technology |
| **IEC** | International Electrotechnical Commission |
| **IoT** | Internet of Things |
| **IP** | Internet Protocol |
| **ISMS** | Information Security Management Systems |
| **ISO** | International Organization for Standardization |
| **ISP** | Internet Service Provider |
| **IT** | Information Technology |
| **MFA** | Multi-Factor Authentication |
| **NFC** | Near Field Communication |
| **QR** | Quick-Response Code |
| **RFID** | Radio-Frequency Identification |
| **URC** | Universal Remote Console |
| **VR** | Virtual Reality |

## 1. What are AAL technologies?

Active Assisted Living (AAL) technologies support older adults and people with disabilities. They help these users live safely and independently in their homes. These technologies include devices, systems, and services that monitor health, assist with daily activities, and enhance safety. They combine advanced technology like sensors, wearables, and smart home systems to provide real-time care and assistance. AAL technologies improve quality of life by addressing health concerns, promoting social engagement, and helping with routine tasks. Their flexibility allows them to adapt to the user's changing needs over time, making them an essential tool for independent living.

One feature of AAL technologies is their ability to monitor health and well-being. For example, many devices continuously track health metrics such as heart rate, blood pressure, and sleep patterns. These data are shared with caregivers or healthcare providers, allowing for early detection of potential health issues. Remote health monitoring allows timely intervention. It also reassures users and their families, reducing the need for frequent in-person check-ups.

In addition to health monitoring, safety is a central focus of AAL technologies. Fall detection systems are a common example, automatically alerting caregivers or emergency services if a fall occurs. This type of system is particularly useful for older adults who may live alone and want the reassurance that help is available if needed. Other safety features like smart locks and motion-activated lights help create a secure, accessible home. They prevent accidents and promote independence.

AAL technologies are also designed to make everyday tasks easier. Smart devices can automate things like lighting, heating, and household appliances, which can be controlled with simple voice commands or apps. For those with mobility issues or cognitive decline, this automation helps maintain their independence and reduces the need for physical effort or remembering complex tasks. For example, a smart stove can be set to turn off automatically if left unattended, preventing kitchen accidents.

Social connection is another important aspect of AAL technologies. They provide tools for staying in touch with family and friends, addressing one of the key challenges for older adults—loneliness. Technologies such as video calls, voice assistants, and social platforms designed for older users help people connect with loved ones without requiring complicated gadgets.

Another benefit of AAL technologies is their potential to reduce healthcare costs. By enabling individuals to manage their health at home, these technologies decrease the need for hospital visits or long-term care facilities. This not only saves money for families but also eases the burden on the healthcare system. Over time, the cost savings can be significant, making AAL technologies a practical investment for everyone involved.

Overall, AAL technologies are essential in enabling older adults and people with disabilities to live independently. They combine health monitoring, safety features, assistance with daily tasks, and tools for social engagement to offer a comprehensive system that adapts to the user's needs.

## 2. What are the most common types of AAL technologies?

Developing AAL technologies is a growing challenge. The goal is to find affordable ways to help people live independently and stay healthy. Beyond making these technologies easy to use, there is a focus on designing devices that support both mental and physical well-being. A wide range of technologies has been created to help people live an active and assisted lifestyle. These technologies can be grouped into several categories:

- **Sensing technologies**: Sensing technologies use sensors to monitor the environment or the person using the technology. These sensors track aspects like movements, activities, or vital signs such as heart rate. Some sensors, like cameras or radar systems, can recognise a person's movements, while others combine signals from multiple sensors for better accuracy. This helps the technology understand the user's behaviour and respond appropriately. Advanced sensing technologies can even predict potential health issues by analysing patterns over time, allowing for early intervention.
- **Mobile technologies**: Smartphones or tablets are valuable tools for gathering information about a person's health and daily activities. They help users manage their health and well-being at home, offering reminders for taking medication or exercising. These technologies use built-in sensors to measure the user's environment or behaviour and suggest activities to promote a healthier lifestyle. Mobile apps can also connect users with healthcare providers or support networks, enhancing communication and access to care.
- **Robotic technologies**: Robotic technologies aim to mimic human abilities to assist users with daily tasks. For example, some robots can recognise hand gestures to help with activities like picking up objects. Others provide companionship, engaging users in conversations and helping with tasks around the house. These robots can reduce feelings of loneliness and assist with maintaining an independent lifestyle.
- **Wearable technologies**: Wearable devices, like fitness trackers or smartwatches, collect and analyse data about a user's activities. They track metrics like steps taken, heart rate, and can even detect falls or other emergencies. However, wearable devices must be energy-efficient to work all day without needing to be charged frequently. Privacy and ease of use are also important concerns, as users want devices that are helpful without feeling invasive. Wearables can be placed on different parts of the body, such as the wrist, to measure energy use while walking or to monitor daily activities. Some advanced wearables can even monitor blood glucose levels or detect irregular heart rhythms, providing valuable health insights.
- **Conversational and gaming technologies**: These technologies use conversation and games to engage users in fun and useful activities. For example, social robots can chat with users and help keep their minds active, while games can encourage people to stay physically and mentally engaged. These activities help promote a healthy, active lifestyle. Interactive platforms can also offer cognitive training exercises, which may help slow cognitive decline or improve mental agility.

## 3. What is the Internet of Things?

The **Internet of Things (IoT)** is a concept where everyday objects, machines, and devices are connected to the Internet. These devices can communicate with each other and share information. The IoT includes anything from your phone and smartwatches to home appliances like fridges and lights, and even industrial machines. The idea is to create a system where devices can work together, making life easier and more efficient.

There are different ways to think about the IoT, depending on what you focus on:

- **Focusing on "Things"**: From this view, the IoT is about objects, or "things," becoming smarter. These objects now have identities, meaning they can interact with each other and with humans. For example, smart home devices like thermostats and security systems can adjust settings or send alerts based on what is happening around them. Here, the focus is on making these things more intelligent and capable, while how they connect is important but secondary.
- **Focusing on the "Internet"**: This perspective looks at how these "things" use the Internet to communicate. Every connected device, whether it is a car or a washing machine, uses the Internet to send and receive data. For example, a fitness tracker sends data about your health to an app on your phone. The challenge here is making sure all these devices can talk to each other smoothly, especially since they may have different ways of processing and storing information.
- **A global view**: This perspective looks at the IoT as a global system of interconnected devices. In this view, the IoT is seen as a vast network where all devices, from small sensors to large machines, can work together. These connected systems create new possibilities, such as smart cities where traffic lights, transportation, and energy systems all communicate to improve efficiency. The IoT connects many different systems, creating a world where devices can collaborate to provide better services.

Although there are several ways to define the IoT, the core idea remains the same: it connects devices to make life more convenient and efficient. Some focus on the devices themselves, while others look at how they communicate or how they fit into a larger system. No matter the approach, by integrating the physical world with the digital one, the IoT opens new possibilities for innovation in various fields, including healthcare, transportation, and home automation.

In the context of AAL, the IoT enhances the quality of life for older adults and people with disabilities. By connecting devices like health monitors, smart home appliances, and wearable sensors, the IoT enables seamless communication between different technologies used in AAL. For example, a wearable health monitor can track vital signs and automatically adjust home settings like temperature or lighting for comfort. It can also send alerts to caregivers or family members if it detects any unusual patterns. This interconnected system allows for real-time responses to the user's needs, making independent living safer and more comfortable.

**4. What role does AI play in AAL technologies?**

Artificial Intelligence (AI) plays a key role in helping older adults or people with disabilities perform everyday tasks without needing to ask for help. Engineers and scientists program these devices with lots of useful information, including details about how people behave, their health data, daily routines, and the specifics of any illnesses. With this data, AI can help the system suggest the best solutions to meet the user's needs, like when to send reminders or alerts.

For example, if a device like an electronic water bottle is powered by AI, it can learn when a person has not met their daily hydration goals. The AI can then decide to send a reminder to encourage the person to drink more water to stay healthy. This is just one example of how AI helps people achieve their goals without needing to constantly monitor themselves.

AI in AAL systems can predict the behaviour of the people who use them, and it helps in several important ways:

- **Helping with daily activities**: AI can assist with activities like moving around the house or going outside. For example, robotic walkers for older adults can help them walk more safely, and smart glasses for visually impaired users can help them navigate their surroundings better.
- **Health monitoring**: AI can keep track of vital signs, like heart rate or blood pressure, to detect any changes that might signal a health issue. This allows caregivers or healthcare professionals to step in before things get worse. AI can even collect this data from people's homes without requiring a hospital visit.
- **Rehabilitation**: AI can also help people recover from injuries or illnesses. For example, if someone is recovering from a fall, AI can suggest the right exercises to speed up their recovery.
- **Facilitating social interaction and mental health support**: AI can help reduce feelings of loneliness by providing virtual companionship through chatbots or voice assistants. These technologies can engage users in conversation, remind them to stay connected with family and friends, or suggest activities to boost their mood. AI-powered apps can also monitor emotional well-being and offer coping strategies or alert caregivers if additional support is needed.
- **Enhancing safety and emergency response**: AI can detect unusual patterns or emergencies, such as a sudden fall or a significant change in daily routines. It can then alert emergency services or designated contacts promptly, ensuring timely assistance.

The main goal of using AI in AAL technologies is to make everyday devices more automatic and accurate. However, the success of these devices depends on the quality of the information used to program them. AI will only be as helpful as the data it is given, so it is important that these systems are matched to the person's specific health conditions or disabilities.

Ethical considerations and data privacy are also important when using AI in AAL technologies. Since these systems often collect personal and health-related information, it is crucial to ensure that data is stored securely and used responsibly. Developers and users must work together to protect privacy and comply with regulations, so people feel confident using these technologies.

## 5. What is the role of social robots in AAL?

As the population ages, there is a growing need for new tools to help older adults and people with disabilities with their daily tasks. Social robots play a significant role in AAL by providing assistance and companionship, helping users live more independently. These robots are designed to interact with humans on a social level, making them more than just functional machines; they also exhibit social behaviours that users expect.

Social robots in AAL can be categorised into two main types:

- **Service robots:** These robots assist with practical tasks to support independent living. They help with fundamental activities like eating, bathing, dressing, and mobility, including navigating around the home. Service robots can manage household chores, maintain safety, monitor individuals who require constant attention, and even fetch items. Their social capabilities make interactions more intuitive, encouraging users to adopt and use them regularly. Research often examines how social functions affect older people's use of these devices in their homes. It also looks at how social features can make the devices easier for them to use.
- **Companion robots:** The primary purpose of companion robots is to provide emotional support and companionship, improving the physical and mental health of older adults. They engage with users through conversation, facial expressions, and gestures, serving as friendly daily companions and intermediaries in peer relationships. Older people tend to view these robots more as companions than as machines, attributing human-like qualities to them. Companion robots can reduce feelings of loneliness and social isolation, common issues associated with ageing. By fostering social participation, they change the role of the older adult from one of passivity to active engagement. Research investigates whether older adults residing in nursing homes might have happier moods when they have companion robots.

Social robots aim to establish a close and effective relationship with a human user. They seek to provide support and achieve measurable improvements in learning, rehabilitation, convalescence, and more. Due to their physical form and multimodal communication channels, social robots can interact socially and benefit users through verbal and nonverbal communication. They use speech, gestures, and even facial expressions to connect with people, making interactions feel more natural. This multimodal communication helps build a strong relationship between the robot and the user. This is crucial for the robot's effectiveness in providing support.

Social robots offer numerous benefits in AAL. By assisting with daily tasks, they enhance independence, enabling older adults to live more independently in their homes. Their companionship can improve mental health by reducing loneliness and depression, promoting a more positive outlook on life. They also increase safety by monitoring the environment for hazards, reminding users to take precautions, and alerting caregivers in emergencies. Furthermore, advanced robots provide personalised interaction, adapting to the user's preferences and routines. This offers tailored support that meets individual needs.

## 6. What is the role of video-based devices in AAL?

Video-based devices, usually cameras, help monitor individuals, keep them safe, and support their health in real-time. These devices are important in AAL systems because they help protect and care for people who may be vulnerable.

One primary use of video-based devices in AAL is **remote monitoring**. Caregivers can keep an eye on older adults without being physically present, using live video feeds or recorded footage to see what is happening at home. This is especially helpful for those who live alone or are at risk of accidents. For instance, video devices can detect falls, a major concern for older people. By analysing movements and identifying signs of a fall, the system can immediately alert caregivers or emergency services, ensuring quick assistance and potentially preventing further injury.

These devices also help in **tracking behaviour and activities over time**. By analysing video data, they can notice changes in behaviour that might indicate health issues or cognitive decline. For example, increased restlessness or decreased activity levels could signal a problem. Unusual actions like forgetting to turn off appliances or wandering at odd hours might be early signs of memory issues. Early detection allows caregivers to adjust care plans, encourage physical activity, schedule medical check-ups, or enhance safety measures.

Video-based devices provide **cognitive and social support** as well. Tools like video calling systems enable seniors to stay connected with family and friends, reducing loneliness, and promoting social engagement. Regular video chats help maintain close relationships, even when loved ones are far away. Interactive video platforms and games can keep the mind active, helping to maintain mental sharpness and delay memory loss or other cognitive problems.

Moreover, these devices can assist with **daily routines and medication management**. Video-equipped systems can guide users through tasks like cooking, exercising, or taking medication by providing visual prompts and demonstrations. This visual assistance helps individuals with memory issues or cognitive impairments complete tasks independently.

**Telemedicine** is another important application. Through video consultations, healthcare professionals can assess patients remotely, offer medical advice, and adjust treatment plans without requiring the person to travel. This is beneficial for those with mobility challenges or living in remote areas, ensuring continuous medical support and early detection of health issues.

In addition, video-based devices enhance **home security**. Cameras can monitor for intruders or suspicious activities, providing peace of mind. Some systems automatically alert authorities or sound alarms if unauthorised entry is detected, adding an extra layer of protection.

Privacy considerations are essential when using video-based devices. To address concerns, many systems focus on movement patterns without capturing identifiable images or using data encryption to protect personal information. Users can control when cameras are active and restrict access, ensuring their privacy is respected.

## 7. What is the role of voice-activated assistants in AAL?

Voice-activated assistants are playing an important role in AAL. They offer a sophisticated but intuitive interface between technology and end users, particularly older individuals or those with disabilities. These systems are powered by advanced speech recognition and artificial intelligence (AI). They enable users to engage with their environment through simple voice commands. They promote a level of accessibility and independence that would otherwise be challenging. Voice assistants integrate with smart home systems, health monitoring devices, and communication platforms. This reduces reliance on complex, direct interactions, empowering users to manage their daily routines with minimal physical effort.

At the core of this innovation is the ability to transform smart homes into truly adaptive environments. Voice-activated assistants empower users to control household appliances, adjust lighting, manage security systems, and organise daily schedules entirely through voice commands. This eliminates the need for manual interaction. This is particularly useful in AAL contexts, where individuals may experience mobility challenges or cognitive impairments. For instance, older adults with limited dexterity or visual impairments can effortlessly modify their surroundings. This is done by issuing simple verbal prompts, significantly enhancing their comfort and reducing potential safety risks. Hands-free access to these systems boosts independence. It reduces reliance on caregivers for routine tasks and fosters a more autonomous lifestyle.

Safety is another crucial dimension where voice-activated assistants make a meaningful contribution. Fall detection, irregular movements, and vital sign changes usually rely on sensors and health devices. Voice assistants play a key role in alerting the users. For instance, if a sensor detects a fall or abnormal health metrics, the voice assistant can immediately prompt the user to confirm their well-being. It can also automatically notify caregivers or emergency services. This integration enhances responsiveness and provides a hands-free method of summoning help. This is especially valuable for those who may struggle to access traditional communication devices during a crisis. Voice-activated assistants offer an extra layer of protection by streamlining communication in emergency scenarios. This contributes to a safer living environment and peace of mind for both users and their families.

The future of voice-activated assistants in AAL is promising with the integration of AI-driven emotional intelligence. As these systems evolve, they are expected to recognise emotional states such as stress, anxiety, or frustration. This will enable a more personalised and responsive care experience. A voice assistant capable of adjusting its interactions based on a user's emotional cues would offer a more comprehensive approach to care. It would address not only the physical but also the psychological well-being of the individual. In addition, machine learning allows these assistants to continuously learn from user interactions, gradually becoming more attuned to individual preferences and behaviours. This ability to anticipate and respond to a user's needs improves the user experience. It makes the technology more seamless and intuitive in daily life in AAL settings.

### 8. What is the role of wearable devices in AAL?

Wearable devices are electronic gadgets worn on the body that monitor and track various health and activity metrics. Common examples include smartwatches, fitness trackers, smart glasses, and wearable heart rate monitors. In the context of AAL, wearable devices play a significant role in enhancing the independence and well-being of older adults and people with disabilities. They assist in tracking physical activity, monitoring health conditions, detecting falls, and supporting individuals with specific disabilities. They make life easier and safer.

One important role of wearable devices is increasing awareness of personal health. These devices can monitor factors such as physical activity levels, dietary habits, heart rate, sleep patterns, and blood pressure. By collecting this data, wearables help individuals understand how their daily habits affect their health. Access to this information enables people to make informed decisions, such as increasing physical activity or adopting healthier eating habits. In essence, wearables support individuals in improving their health and making positive changes in their everyday lives.

Wearable devices can also be integrated with electronic health records, enhancing health management. Wearables collect vital data for people with chronic conditions, like hypertension or diabetes. This includes activity levels, glucose readings, and blood pressure. This data helps healthcare providers create better treatment plans, as it is accurate and obtained in real-time. Continuous monitoring can lead to early detection of potential health issues, allowing for timely interventions.

Another key role of wearables is providing a sense of security. These devices can alert the user or others in case of falls, disorientation, or medical emergencies like heart problems. Knowing that assistance can be summoned quickly provides peace of mind. For example, if an individual gets lost or experiences a health issue, the wearable device can help locate them or detect the problem promptly. This added sense of security enables older adults to live more confidently on their own, allowing them to remain in their homes longer while knowing that help is available if needed.

Wearables are also beneficial for people with specific disabilities. For individuals with hearing impairments, wearable devices can connect to smartphones to enhance sound quality and provide alerts through vibrations or visual signals. Wearables can assist those with visual impairments by reading text aloud, describing scenes or objects, and helping them navigate their surroundings. This technology empowers people to access information more easily and navigate their environment with greater confidence, making daily activities more comfortable.

In addition to these roles, wearable devices contribute to social engagement and mental well-being. Some wearables can track mood, stress levels, and sleep quality, offering insights that help manage mental health. They can also facilitate communication by sending reminders to connect with friends and family or participate in social activities. By promoting both physical and mental health, wearables support an integrated approach to well-being.

### 9. How do AAL technologies benefit older adults?

AAL technologies provide many benefits to older adults, helping them live more independently, enhancing their safety, and offering remote health monitoring. Here are different ways AAL technologies benefit older adults:

1. **Providing companionship:** Virtual assistants and communication platforms, can help reduce feelings of loneliness and isolation. These technologies enable older adults to stay connected with family and friends through video calls, messaging, and social media. Virtual assistants can engage users in conversation, provide entertainment, and offer reminders, helping individuals feel less alone, especially if they live by themselves or have limited mobility. They can also provide emotional support and reduce stress.
2. **Assisting with daily tasks:** Smart home devices can automate lighting, heating, and appliances, making it easier for older adults to manage their homes. For example, smart refrigerators can monitor food supplies and suggest grocery lists, while automated pill dispensers remind users to take their medication or attend appointments. Wearable devices can assist with mobility by tracking movement and providing navigation assistance, making everyday life more convenient.
3. **Enhancing safety:** Devices like motion sensors and smart home systems can detect unusual activities, such as prolonged inactivity or wandering, and alert caregivers or emergency services. Wearable devices monitor vital signs and detect falls, automatically sending alerts if assistance is needed. Smart locks and security cameras enhance home security by controlling access and monitoring the surroundings. By providing immediate responses to emergencies and preventing accidents, these technologies allow older adults to live independently with more confidence.
4. **Cognitive stimulation and therapy:** Brain-training apps and interactive games engage users in activities that keep their minds active. These activities might include puzzles, memory games, or learning new skills through online courses. Some devices are used in physical or occupational therapy, helping to make exercises more enjoyable and effective. Virtual reality experiences can provide stimulating environments for relaxation or cognitive engagement, helping to maintain mental sharpness and overall well-being.
5. **Convenient shopping and services:** Smart devices can use online platforms to order groceries, medications, and other necessities without leaving home. Ride-sharing apps help them get around without needing to drive, offering a more convenient and safe way to travel. Voice-activated assistants simplify tasks like setting reminders, making appointments, or controlling home appliances.
6. **Reducing caregiver burden:** AAL technologies help reduce the workload for caregivers by offering automated support for daily tasks and health monitoring. This allows older adults to live independently for a longer time in their own homes, reducing the need for institutional care. The support provided by these technologies offers peace of mind for both caregivers and older adults, knowing that assistance is available when needed. Caregivers can focus on providing emotional support rather than managing every aspect of daily care.

## 10. How do AAL technologies assist people with disabilities?

AAL technologies help people with disabilities live more independently, stay safe, and monitor their health remotely. These technologies offer various functions and can be customised to fit specific needs. Here are ten ways AAL technologies support people with disabilities:

1. **Increasing independence:** AAL technologies enable people with disabilities to perform daily tasks independently, reducing their need for caregivers. For example, smart home devices allow users to control lighting, temperature, and appliances with voice commands or simple controls. Everyday tasks like locking doors or turning off the stove can also be automated, making life easier.
2. **Enhancing safety:** Service robots, cameras, and sensors, monitor for falls, unusual behaviour, or health emergencies. They automatically alert caregivers or emergency services when needed. Additional features, like surveillance cameras and smart locks, help create a more secure environment.
3. **Health monitoring:** Wearable devices, such as smartwatches, track vital signs like heart rate, blood pressure, and blood sugar levels. If something seems wrong, the system can notify caregivers, allowing for quick action when needed.
4. **Improving quality of life:** AAL systems make life more convenient for people with disabilities. For example, instead of needing frequent doctor visits, users can monitor their health from home. This is particularly helpful for people with mobility issues or those who find it difficult to travel.
5. **Providing reminders:** AAL devices can remind users to take medication, attend appointments, or complete daily tasks. These systems also provide step-by-step instructions for activities, making it easier to manage daily routines.
6. **Assisting communication:** AAL technologies can help people with speech or hearing impairments communicate more easily. Screen readers, eye-tracking technology, and speech-to-text apps make it easier to connect with family, friends, and healthcare providers, reducing feelings of social isolation.
7. **Customising interactions:** AAL systems can be programmed to suit individual needs, such as using simple language or responding to specific gestures. This makes the technology easier to use and more effective for each person.
8. **Improving mobility:** Smart wheelchairs and smart glasses help users move around more efficiently by avoiding obstacles and navigating tough terrain. These tools give people more freedom to move independently.
9. **Tailoring solutions to specific disabilities:** AAL devices can be customised to meet user's specific needs. For example, visual aids can help people with vision impairments, while mobility aids can assist those with physical limitations.
10. **Providing peace of mind:** AAL technologies give both users and their families peace of mind. Knowing that health is being monitored reduces stress for caregivers and can lower overall healthcare costs.

**Use Case Scenario 1: Balancing independence and care. Teresa's AAL journey**

Teresa, 74, lives with her 87-year-old husband in a third-floor apartment without a lift in Porto, Portugal. While she is active in her community and enjoys her independence, caring for her husband, who has multiple health issues, is becoming increasingly demanding. The added difficulty of climbing stairs is also taking a toll on her. Teresa is considering moving to a more accessible apartment to ease the burden while ensuring her husband's well-being, but she wants to maintain her active lifestyle.

**Health concerns and challenges**

Teresa's husband's health has been in decline due to his chronic conditions, requiring constant care and attention. She helps him with hygiene, getting dressed, and managing his medications, but the strain of caregiving is growing. The increasing difficulty of climbing stairs has also started to take a toll on Teresa, who experiences joint pain and is concerned about her future mobility.

Teresa is also fearful of her own health declining, which would make it impossible for her to care for her husband or maintain her active routine. She is considering a move to a more accessible apartment but is worried about the emotional and physical stress involved in relocating.

**Technology solutions for their needs**

The couple's situation presents an opportunity to adopt AAL technologies to support both Teresa's independence and her husband's safety. These technologies can play a pivotal role in addressing Teresa's needs for mobility, caregiving support, and safety while allowing her to maintain her lifestyle.

**1. Fall detection and emergency alert systems:** Given Teresa's husband's limited mobility and increased physical dependence, installing an emergency alert system with fall detection features in their current apartment would be beneficial. These systems can be discreet and non-intrusive, using wearable devices or sensor-based monitors in high-risk areas like the bathroom. This will help prevent accidents and ensure immediate assistance is available if her husband falls, providing peace of mind for Teresa when she is out of the house.

**2. Smart home modifications:** To assist with Teresa's growing difficulty climbing stairs and her husband's limited mobility, smart home modifications could help. Stairlifts could be installed in their current apartment to make it easier for Teresa to access the upper floors. Additionally, automated lighting systems and voice-activated controls for household appliances can reduce the physical strain of managing the home. Teresa could control lights, curtains, and even kitchen appliances through simple voice commands, making her daily activities less taxing.

**3. Health monitoring devices for Teresa's husband:** Teresa's husband could benefit from wearable health monitoring devices that track vital signs like heart rate, blood pressure, and oxygen levels. These devices can send real-time data to healthcare professionals or alert Teresa and their son if any concerning changes arise. This proactive monitoring can help manage his chronic conditions and ensure that any medical issues are detected early, reducing the need for frequent medical appointments.

**4. Remote caregiver communication:** To ensure that Teresa's husband receives appropriate care when she is out of the house, a remote monitoring system can be integrated. The system would allow Teresa to check in on her husband via her mobile device, ensuring that he is safe and comfortable. This system can also provide updates and alerts to their son, who lives nearby, ensuring that the family remains connected and informed.

**5. Mobility solutions for Teresa:** To help maintain her mobility and reduce the impact of joint pain, Teresa can adopt wearable devices such as smart insoles that track her steps and posture. These devices can offer real-time feedback to prevent strain and injury, helping her stay active without overexerting herself. Furthermore, if she chooses to stay in her current apartment, portable mobility aids like canes with built-in alarms could also help her navigate the stairs safely.

**Overcoming Challenges and Fears**

Despite the benefits of adopting AAL technologies, Teresa has several concerns that need addressing.

**1. Fear of isolation and dependency on technology:** While Teresa is eager to maintain her independence, she is concerned that too much reliance on technology may distance her from her husband and reduce the human touch in caregiving. To alleviate this concern, Teresa can combine AAL technologies with her existing care routines, ensuring that she still spends quality time with her husband while using the technology as a safety net. For instance, remote health monitoring allows her to step away for her classes without feeling disconnected.

**2. Concern about moving to a new home:** Teresa is reluctant to leave her current apartment and the familiarity of her neighbourhood, but the physical challenges of climbing stairs daily are becoming untenable. An accessible apartment with fewer stairs and better support for mobility could improve their quality of life. The transition may be difficult emotionally, but by choosing a location within their current neighbourhood, Teresa can still maintain her social connections. She can also involve her son in the process to ensure the move is as smooth as possible.

**3. Privacy concerns regarding monitoring systems:** Both Teresa and her husband might have concerns about privacy when it comes to installing cameras or wearable monitoring devices. To address these concerns, the systems should be chosen carefully, prioritizing non-intrusive options like motion sensors rather than cameras. Transparency with Teresa's husband about how the technology will be used can also help ease any discomfort he may have.

**Conclusion**

For Teresa, AAL technologies offer a way to maintain her active lifestyle while ensuring her husband's safety. Devices like fall detection systems, health monitors, and smart home automation can provide support in their current home. However, as her husband's needs increase, Teresa might consider moving to a more accessible apartment to ease the physical strain. By choosing technologies that suit their privacy and daily routines, they can continue living independently and comfortably.

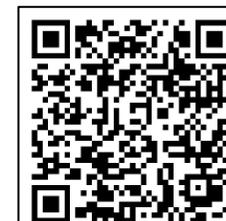
Learn more about Teresa

## 11. How do AAL technologies assist people with cognitive decline and dementia?

People with cognitive decline, such as dementia, often face challenges in performing everyday tasks due to memory loss and reduced cognitive abilities. AAL technologies offer valuable support by making daily life easier, enhancing safety, and improving the quality of care for both individuals with dementia and their caregivers. Here are several ways AAL technologies assist people with cognitive decline:

1. **Passive sensor devices:** They operate automatically without requiring any action from the user, making them ideal for individuals with memory impairments. They enhance safety by monitoring the environment and alerting caregivers when necessary. For example, stove guards monitor the stove and detect if it is left on too long or overheating; they can send alerts to caregivers or automatically shut off the stove to prevent fires. Bed sensors detect when a person gets out of bed at night and alert caregivers if they do not return within a set time. Motion sensors and lights identify movement or inactivity patterns, turning on lights to prevent falls or alerting caregivers if unusual inactivity occurs. Door sensors send alerts when doors or windows are opened, helping caregivers monitor if the person leaves the room or house, which is crucial for preventing wandering and ensuring safety.
2. **Tracking devices:** GPS trackers enhance safety and independence by allowing caregivers to know the person's whereabouts. Worn as wristbands or pendants, they help prevent dangerous situations like wandering away from home. If the individual becomes lost or disoriented, caregivers can quickly locate them, providing peace of mind.
3. **Everyday life technologies:** AAL technologies assist with daily living by providing reminders and simplifying tasks, improving memory, independence, and overall quality of life. Electronic medication dispensers organise medications and provide reminders when it is time to take them, sometimes only releasing the correct dose at the scheduled time to prevent errors. Digital calendars or watches offer memory aids by displaying appointments and tasks, sending alerts to help individuals stay organised. Robotic pets and companions provide companionship and emotional support, responding to touch and sound without the responsibilities of caring for a live pet.
4. **Communication technology:** Maintaining social connections is vital for emotional well-being. User-friendly devices enable individuals with cognitive decline to stay connected with family and friends. Simplified computers and tablets with intuitive interfaces allow users to make video calls, send messages, and share photos easily. Voice-activated assistants respond to voice commands, helping with calls, reminders, or accessing information without navigating complex menus.

While AAL technologies offer many benefits, it is important to choose devices that match the individual's abilities and needs. Some technologies may be challenging for those with advanced dementia. For example, emergency call buttons may not be effective if the person forgets how to use them, potentially creating a false sense of security. Therefore, passive systems that do not rely on user initiation are often more reliable.

**12. How can AAL technologies aid in fall prevention?**

AAL technologies offer several solutions to prevent and detect falls, helping older adults and people with disabilities stay safer and more independent. Falls are a significant concern for this population, as they can lead to serious injuries, reduced mobility, and a loss of confidence in daily activities. Fall detection systems use different methods, each with its own benefits and challenges, to provide security and peace of mind for users and caregivers:

1. **Active sensor technology:** Active sensors require the user to push a button or take some action in case of a fall or emergency. For instance, social alarms, often worn as a pendant or wristband. After a fall, the user presses the alarm button to send an alert to a caregiver, family member, or monitoring centre. This alert can trigger a response, such as a phone call to check on the user or dispatch emergency services if needed. Most social alarms are connected to a monitoring system in the home, but some also work outside the house, using GPS to track the user's location. This feature is particularly helpful for those who may wander or spend time outdoors. However, active sensors rely on the user's ability to recognise the emergency and operate the device, which can be challenging for people with cognitive impairments or in situations where the user is unconscious or unable to move.
2. **Passive sensor technology:** Passive sensors automatically detect falls or unusual activity and send alerts without requiring any action from the user. This makes them especially useful for individuals who may not be able to press an alarm in an emergency. Passive sensors include:

- **Acceleration sensors:** These sensors, worn on the body as a watch or necklace, automatically detect falls and send alerts to caregivers. They work best in cases of quick, sudden falls but may not pick up slower movements like sliding from a chair or bed.
- **Camera-based radar or infrared motion sensors:** These sensors monitor the person's movements and can detect when they leave their bed or room and do not return, fall, or call for help. The system then sends an alert with an anonymous image to the caregiver's mobile app, allowing them to assess the situation without invading privacy. Radar sensors can be placed anywhere in the home, while infrared sensors may have some limitations.
- **Pressure-based sensors: Smart tiles** and other pressure sensors are placed on the floor to detect falls and send alerts. They are less intrusive than wearable devices or cameras and are often used in homes or nursing facilities. **Instrumented insoles** are another form of pressure-based sensors. They monitor a person's balance and walking patterns, helping assess fall risk, and can be worn anywhere.
3. **Advanced fall prevention systems:** Some AAL technologies combine multiple sensors and data sources to provide comprehensive fall prevention. These systems integrate wearable devices, environmental sensors, smart home technology, and artificial intelligence to monitor movements, detect hazards, and predict falls. By analysing patterns, they can identify increased fall risk and prompt interventions, such as suggesting rest or alerting caregivers.

## 13. How can AAL technologies aid in mobility and transportation?

Older adults often avoid going outside because they feel unsafe, which can lead to health problems like muscle loss, osteoporosis, arthritis, heart disease, depression, and cognitive decline. AAL technologies can help older adults and people with physical disabilities stay more active and mobile, giving them the confidence to move around while ensuring their safety. These technologies focus on providing support for travel, navigation, and safety, making transportation easier and safer for users.

One important way AAL technologies help is by offering **travel planning and navigation** tools. These applications allow older adults to use public transportation and navigate unfamiliar places with more confidence. They provide detailed directions, real-time updates, and safety features for travelling. For example, navigation apps can show the direction and distance to the next bus stop, provide schedules for buses or trains, and highlight important landmarks along the route. Some apps offer voice-guided navigation, which can be especially helpful for users with visual impairment. These tools help users feel secure when they travel alone or in unknown places. They can also suggest accessible routes that accommodate wheelchairs or walkers, ensuring users can reach their destinations without obstacles.

Another key feature is **enhancing outdoor safety**. AAL technologies can detect potential dangers like uneven sidewalks, steep inclines, or obstacles, guiding users to safe, accessible walking or driving routes. Wearable devices with sensors can monitor the user's surroundings and provide alerts if hazards are detected. For instance, a smart cane equipped with sensors can vibrate to warn of obstacles ahead. If a person becomes disoriented, the system can share their location with a caregiver or relative, allowing them to receive support when needed. This feature provides users with a greater sense of freedom while knowing help is available if they get lost or face challenges. It also reduces anxiety for both the users and their families, promoting more frequent and confident outings.

**Locating and emergency alerting** are also essential for improving mobility and transportation. These systems can track a person's location and alert caregivers if something unusual happens, like the person leaving a designated safe area or not returning home within a set time. GPS trackers can be worn as wristbands or pendants, and some can even detect if the user has fallen or is immobile for an unusual period. If the user cannot press an alarm, such as in the case of dementia or Down syndrome, these technologies can automatically detect irregular behaviour or lack of movement and notify caregivers. This extra layer of security gives the user more freedom while keeping them safe.

AAL technologies also offer **mobility assistance** for people with physical limitations. Smart wheelchairs and walkers help users move around more easily and safely. These devices can have features like obstacle detection, automatic braking, and navigation assistance. For example, a smart wheelchair might have sensors that prevent it from colliding with objects or tipping over uneven surfaces. Some can even be controlled by voice commands or slight movements, aiding users with limited mobility in their hands or arms. These devices provide direct support for movement, making everyday activities more manageable. They enable users to participate in social events, run errands, or simply enjoy time outdoors.

### 14. How can AAL technologies enhance home safety and security?

AAL technologies help improve home safety by detecting risks early and ensuring the safety of older adults and people with disabilities. These technologies monitor daily activities, detect emergencies, and provide real-time alerts to caregivers or family members, offering peace of mind. AAL systems provide continuous monitoring and emergency detection, making homes safer and more secure.

A key part of AAL home security is **continuous monitoring**, which covers both general and safety-related aspects:

- **Monitoring daily activities**: This includes tracking movements, meals, and medication intake. If the person does not move for an unusually long time, caregivers or family members are notified.
- **Safety monitoring**: The system keeps an eye on various risks around the home, including:
  1. Nighttime movement. For example, a light automatically turns on when the person leaves their bed, helping prevent falls on the way to the bathroom.
  2. Temperature monitoring. The system detects if it is too hot or cold inside the home.
  3. Monitoring doors and windows. Alerts are sent if doors or windows are left open at night. Family members can remotely lock doors or grant access to neighbours or emergency services.
  4. Detecting smoke, fire, and gas leaks. For instance, if the stove has been on for too long, the system will automatically shut it off.
  5. Water leak detection. Alerts are triggered if a tap is left running or there is a risk of water overflow.
  6. Managing risky appliances. The system can automatically turn off household appliances, such as a stove, if it has been forgotten or left on while the house is empty.

By analysing data from these monitored objects, the system identifies any unusual behaviour or risks and sends alerts automatically to ensure quick action can be taken.

AAL technologies also improve **comfort and convenience** at home with:

- **Temperature control**: The system can pre-set desired temperatures for each room. For example, it heats the bathroom early in the morning or evening.
- **Automatic lighting**: To save energy and increase security, lights turn on when there's movement and switch off when no one is present.
- **Air quality monitoring**: The system measures humidity and $CO_2$ levels to improve air quality and reduce the risk of illness.

In case of an emergency, AAL technologies are equipped with **emergency detection** features:

- They detect falls, extended periods of inactivity, or if the person leaves home unexpectedly.
- The system can distinguish between safe and unsafe situations and automatically trigger mobile safety alarms to notify caregivers or family members.
- In dangerous situations, such as fires or gas leaks, the system can help evacuate people from the home.

## 15. How can AAL technologies help with medication management?

AAL technologies help people manage medication effectively. Using wireless technologies like Bluetooth, Near Field Communication (NFC), and Radio-Frequency Identification (RFID), these systems—ranging from mobile apps to wearable devices—send reminders and alerts to ensure that people take their medication on time. This technology can also notify users when it is time to reorder their prescriptions, helping to avoid missed doses.

One benefit of AAL technologies is the ability to automate medication dispensing. Devices like smart pillboxes can be pre-loaded with medicine for a week or a month, dispensing the correct dosage at the scheduled time. This automation reduces the chances of human error, ensuring the right dose is taken. In cases where a dose is missed, these devices can alert family members or caregivers. These systems can also track the remaining amount of medicine and provide reminders to refill prescriptions when supplies run low.

Some AAL technologies can connect with healthcare providers. They allow remote monitoring of medication adherence. They can track daily activity, medication intake, and overall health data, sharing this information with doctors or caregivers. This helps ensure that the patient is following their treatment plan without needing frequent in-person visits.

Here are some examples of how AAL technologies help with medication management:

1. **Reading prescription labels aloud**: For individuals with visual impairments or reading difficulties, AAL technologies can assist by reading prescription labels aloud. Using RFID text-to-speech technology, a microchip embedded in the prescription bottle stores the label and leaflet data. When activated, the information is audibly read to the individual, ensuring they understand dosage instructions and warnings, which enhances safety and compliance
2. **Monitoring glucose levels in real-time**: Wearable sensors can continuously track glucose levels for individuals with diabetes. These devices send the data to a smart device or directly to healthcare providers, making it easier to manage the condition. Real-time monitoring allows for immediate adjustments in diet, activity, or medication, improving overall glucose control and reducing the risk of complications.
3. **Finding medication**: It is not uncommon for individuals to misplace their medications. Bluetooth tags attached to medication packages can help locate the medicine through a smartphone app. This ensures that patients do not miss doses due to misplaced medications, promoting adherence to their treatment plan.
4. **Medication management apps**: Mobile applications can provide personalised medication schedules, reminders, and educational information about the medications being taken. Some apps allow users to log their medication intake, note any side effects, and set up alerts for upcoming doses. These apps can also generate reports that can be shared with healthcare providers during appointments.

## 16. How can AAL technologies promote social connectedness?

As people age, they face physical challenges, but their social environment also plays a crucial role in their health and well-being. Staying socially engaged is vital for mental health, as social isolation can lead to loneliness and depression.

Different technologies have been created to improve social interaction and create a sense of connectedness. One example is **social networking applications** tailored for older adults. These platforms allow users to connect with family members, friends, and caregivers. They gather user information to match people with similar interests, facilitating the sharing of information and communication. Within these networks, **recommender systems** suggest new people to connect with, activities to participate in, and events to attend based on shared interests and health conditions. This helps raise awareness of opportunities for social interaction and fosters a sense of community.

These social networks are often accessed through user-friendly devices like tablets or specially designed interfaces that provide easy access to the system. As users interact with the platform, their profiles are updated to reflect their activities and preferences. Some technologies can also track users' mobility or engagement levels, adding helpful information to the social network. This makes it easier to adapt recommendations to their physical capabilities and interests, ensuring that suggested activities are appropriate and accessible.

**Conversational technologies** are another useful tool for promoting social connectedness. Advances in natural language processing have made it possible to create chatbots and virtual assistants that engage users in meaningful conversations. These digital companions can provide social interaction, emotional support, and even cognitive stimulation. By engaging in regular dialogues, they help reduce feelings of loneliness and keep the mind active. Some are designed to detect changes in the user's mood or emotional state, offering supportive responses or suggesting activities that might improve their well-being.

In addition to one-on-one interactions, AAL technologies can facilitate **group activities and virtual gatherings**. Video conferencing tools and online platforms enable older adults to participate in social events, support groups, and educational classes from the comfort of their homes. This is especially beneficial for those with mobility limitations or health conditions that make it difficult to attend in-person events. Virtual reality environments can also provide immersive experiences where users can interact with others in simulated settings, such as attending a virtual concert or exploring a virtual museum together.

AAL technologies also support **intergenerational connections**. Platforms can facilitate communication between older adults and younger family members, promoting the exchange of knowledge and experiences. This not only strengthens family bonds but also helps older adults stay engaged with the changing world.

To ensure that these technologies are effective, it is important that they are **accessible and easy to use**. User interfaces should be designed with simplicity in mind, featuring large icons, clear text, and intuitive navigation. Providing training and support can also help older adults become comfortable with new technologies, increasing their willingness to use them regularly.

## 17. How do AAL technologies aid in emergency situations?

AAL technologies play a crucial role in managing emergencies for older adults and people with disabilities. By enhancing safety and providing immediate assistance when needed, these technologies monitor risks such as fires, gas leaks, burglaries, falls, sudden health issues, and wandering. They ensure timely alerts to caregivers, family members, or emergency services.

AAL devices often include sensors and camera systems installed in rooms to enhance safety and monitoring. These technologies are particularly helpful for families with members who have cognitive impairments, such as dementia, or physical limitations due to age. For example, sensors can detect falls or alert if the person has left the house unexpectedly. When such an event occurs, an alert is sent directly to a smartphone, allowing a family member to check the camera feed to determine if assistance is needed. This technology enables alarms to be sent promptly to a relative or private assistant. However, these sensors may not be suitable for households with pets, as animals can accidentally trigger the fall detection sensors. In such cases, alternative technologies that differentiate between human and pet movements may be considered to avoid false alarms.

Security and protection systems are key components of AAL technologies in emergencies. Environmental sensors detect hazards like smoke, gas leaks, or significant temperature changes, alerting occupants and emergency services promptly to prevent harm. By continuously monitoring the home's environment, they provide an added layer of safety against common household dangers. Fall detection sensors identify when a person has fallen and automatically send alerts, which is crucial for individuals who may not be able to call for help themselves. Some systems use accelerometers and gyroscopes in wearable devices, while others employ floor sensors or cameras to detect falls. Motion sensors and door alarms monitor movement within the home. They alert caregivers if an individual with cognitive impairments leaves unexpectedly or exhibits unusual activity. This helps prevent dangerous situations like wandering, which can lead to injury or getting lost.

Wearable emergency devices also play a significant role. Health monitoring wearables, such as smartwatches or pendants, measure vital signs like heart rate, breathing rate, and blood pressure. If these measurements fall outside normal ranges, they alert healthcare workers or designated contacts for prompt intervention. This continuous monitoring is ideal for individuals who prefer privacy but still need their health closely observed. GPS localisation systems are wearable devices with tracking capabilities that monitor the location of individuals with cognitive limitations. Caregivers are notified if the person leaves a safe area or does not return home on time, allowing for quick assistance and ensuring their safety.

By integrating these technologies into daily life, AAL systems provide multiple ways to manage emergencies. Immediate alerts ensure help arrives promptly during crises, reducing the risk of complications from delayed assistance. These technologies provide peace of mind. They assure people that, in emergencies, help will come quickly. This enhances everyone's well-being. Additionally, they enable older adults and people with disabilities to live confidently, knowing assistance is available if needed, supporting their desire to maintain autonomy.

## 18. How do AAL technologies facilitate remote healthcare?

AAL technologies play an essential role in improving healthcare by allowing people to access medical services from home. Using advanced tools like **artificial intelligence (AI)**, the **Internet of Things (IoT)**, and **telemedicine**, AAL technologies make healthcare more accessible and efficient. They support healthcare monitoring, communication, and management, especially for older adults, people with disabilities, and those in remote areas. These technologies enable patients to receive care from the comfort of their homes, leading to better health outcomes and increased satisfaction.

One major way AAL supports remote healthcare is through **telehealth platforms**. These systems enable users to schedule virtual consultations with doctors, therapists, or specialists. Through video calls, patients can receive medical advice, mental health support, or check-ups without needing to travel to a clinic or hospital. This makes it much easier for people to access the care they need.

**Remote health monitoring** is another important feature of AAL technologies. Wearable devices and sensors continuously track vital signs like heart rate, blood pressure, glucose levels, and sleep patterns. This data is sent to healthcare providers in real-time, allowing them to monitor patients' health remotely. If something unusual is detected, doctors can intervene before the condition worsens.

For people with **chronic diseases** like diabetes, high blood pressure, or respiratory conditions, AAL technologies provide tools for continuous monitoring and self-care. These devices track symptoms, remind users to take their medication, and send updates to healthcare providers. This approach helps patients manage their conditions without frequent in-person visits. It also ensures that everyone involved in the patient's care has up-to-date information, improving treatment decisions and care management.

In cases of **emergencies**, many AAL systems are equipped with alert features. If a patient falls, suddenly declines, or has a medical crisis, the system can alert emergency services or healthcare providers. Some systems also offer real-time location tracking or provide detailed health data to responders, further improving the chances of timely and effective intervention. This quick response can be life-saving, reducing risks during critical situations.

AAL technologies also support **rehabilitation and therapy** at home. Virtual reality (VR) tools or interactive platforms guide users through physical or occupational therapy exercises. These exercises are often supervised by healthcare professionals, ensuring that users follow the correct movements and make progress without needing to visit a clinic. Additionally, AAL systems can track recovery data over time, allowing therapists to adjust treatment plans based on real-time progress.

Another advantage of AAL technologies is their **cost-effectiveness**. These systems save money for patients and providers. They do this by reducing the need for in-person visits, cutting transport costs, and lowering hospital readmission rates. They also enable proactive healthcare. This prevents complications that could require costly treatments.

## 19. How do AAL technologies preserve user independence?

AAL technologies help older adults and people with disabilities live independently. These technologies use smart devices, sensors, and communication systems in everyday settings. They allow users to live safely in their own homes without constant supervision or the need to move to care facilities.

One way AAL technologies support independence is by helping with daily tasks. Smart home systems can automate things like controlling lights, heating, and appliances. This makes it easier for users to manage their homes without physical strain. Voice-activated assistants let users operate devices hands-free, which is especially helpful for those with mobility challenges. For example, a person can adjust the thermostat or turn off lights without moving around the house, reducing the risk of accidents.

AAL technologies also improve safety, which is crucial for independent living. Devices like fall detectors, wearable emergency alerts, and environmental monitors can detect emergencies such as falls, fires, or gas leaks. They automatically notify caregivers or emergency services. This quick response gives users the confidence to live alone, knowing that help is available if needed. For instance, if a fall is detected, the system can alert a family member or medical professional for prompt assistance.

Health monitoring is another important aspect that supports independence. Wearable devices can track vital signs like heart rate, blood pressure, and blood sugar levels. This allows users to manage health conditions without frequent hospital visits. The data can be shared remotely with healthcare professionals, who can intervene or adjust treatments as needed. This proactive approach empowers users to control their health, reducing dependence on others.

For people with memory problems, cognitive support technologies help with daily functioning. Tools like electronic medication reminders, digital calendars, and prompt systems help users stick to their routines. By providing gentle reminders for tasks like taking medicine or attending appointments, these technologies reduce reliance on caregivers for daily activities, allowing users to maintain independence.

AAL technologies also promote social connections, which are vital for mental health. Communication platforms and social networking apps enable users to stay in touch with family and friends, join community events, and access support networks. This engagement reduces feelings of isolation. For example, video calling allows users to connect with loved ones easily, strengthening relationships and encouraging active participation in social life.

Improving mobility is another way AAL technologies preserve independence. Mobility aids like smart wheelchairs, robotic walkers, and navigation apps help users move around more easily and safely. These devices can assist with balance, provide directions, and detect obstacles, enabling users to navigate their surroundings confidently. By overcoming physical barriers, users can stay active, which contributes to their independence and quality of life.

## 20. How do AAL technologies support chronic disease management?

AAL technologies help people, especially older adults, manage chronic diseases more easily. They allow individuals to live comfortably and independently by continuously monitoring health, offering personalised care, and providing remote assistance. These tools play a vital role in managing chronic conditions by keeping track of health, providing customised support, enabling remote consultations with doctors, simplifying daily tasks, and offering emotional support.

A key feature of AAL technologies is continuous health monitoring. These tools use sensors to track vital signs like heart rate, blood pressure, blood sugar, and oxygen levels. If something unusual is detected, doctors or caregivers are alerted immediately, allowing for quick action. This helps prevent small issues from becoming bigger problems. For example, if a person's blood sugar drops too low, the system can notify a family member or healthcare provider to help right away.

AAL technologies also offer personalised care. Since everyone's health needs are different, these systems learn what each person requires. They can remind individuals to take their medication at the right time, follow a healthy diet, or do exercises recommended by their doctor. This tailored approach ensures that the care plan fits the individual's specific condition, making treatments more effective.

With remote health monitoring, AAL technologies make it easier for people to receive medical attention without leaving their homes. Sensors in the home or wearable devices send health information directly to doctors and nurses. This is especially helpful for those who live far from healthcare facilities or have difficulty travelling. Video calls with doctors allow for regular check-ups and consultations, reducing the need for in-person visits and lowering the risk of exposure to illnesses.

AAL technologies also assist with daily activities, which can become more challenging for people with chronic diseases. Smart devices can control lighting, temperature, and home security systems through voice commands or automatic settings. This makes daily tasks easier and helps individuals feel safer and more comfortable in their homes, especially if they have limited mobility. For example, smart appliances can make cooking simpler, and automated reminders can help with scheduling appointments.

In addition to physical support, AAL technologies provide emotional and social support. Living with a chronic disease can sometimes feel isolating and stressful. Some AAL systems include virtual companions that can engage in conversation and provide company. These systems also make it easier to stay connected with family and friends through video calls or participate in online support groups. Social interaction is important for mental health, and these technologies help reduce feelings of loneliness.

Furthermore, AAL technologies can assist with rehabilitation and therapy. Interactive applications and devices can guide users through physical therapy exercises, monitor progress, and adjust programs as needed. This supports ongoing treatment plans and helps individuals recover more effectively.

## Use Case Scenario 2: Carmen's Experience with AAL Technologies in a Care Home

Carmen, an 87-year-old care home resident in suburban Spain, enjoys connecting with the community but struggles with chronic health conditions like arthritis, hypertension, and osteoporosis. These issues have led to frequent nighttime falls, causing injuries and long recovery times. Her growing anxiety about safety has limited her participation in social activities, impacting her emotional well-being and increasing her sense of isolation.

The care home plans to install video monitoring in Carmen's room to detect falls and alert staff quickly. While she understands the benefits, Carmen worries that staff will rely too much on the technology, reducing personal interactions. She fears this could leave her feeling abandoned and neglected.

**Carmen's need for safety and emotional support**

Carmen's primary concern is safety, especially when moving around at night. The video monitoring system aims to address this by providing continuous surveillance, ensuring that any fall or risky situation is immediately detected, and allowing staff to respond quickly. This technology can prevent injuries by enabling faster intervention during emergencies. Given Carmen's history of falls, this is a critical step toward safeguarding her health.

However, Carmen's emotional needs are equally important. Her sense of belonging and connection with the care home staff and fellow residents helps her maintain a positive outlook despite her physical limitations. Carmen enjoys engaging in activities like sharing meals with other residents, playing bingo before lunch, and participating in social gatherings. The fear that the new video system might reduce personal interactions and make her feel isolated is a significant concern for her.

**Addressing Carmen's fears about the technology**

The care home management must acknowledge and address Carmen's concerns about feeling abandoned. It is important to involve her in the decision-making process and explain how the video monitoring system works. Carmen should be reassured that the technology is there to complement, not replace, the personal care and attention she receives from staff.

A key strategy would be to establish a clear protocol that combines technology with personal care. For instance, caregivers could make regular visits to check on Carmen throughout the day, ensuring that she still receives the personal attention she values. Even with the video monitoring in place, staff can continue to engage with her and maintain the relationships that bring her comfort. This way, Carmen will feel reassured that the technology is not meant to reduce her interactions with caregivers but rather to enhance her safety.

**Enhancing Carmen's sense of community and participation**

In addition to addressing her fears, it is crucial to keep Carmen actively engaged in the care home community despite her physical limitations. The care home can create opportunities for Carmen to participate in activities that accommodate her needs. For example, activities such as seated exercises, storytelling sessions, or arts and crafts could be adapted to include residents with mobility issues like Carmen. This would allow her to remain involved in social activities and maintain her sense of connection with other residents.

Staff should also encourage peer support within the care home community. Since Carmen already has emotional support from some of her fellow residents, fostering this network can help her feel less isolated. Organising group activities where residents can interact and support each other in a relaxed and informal setting could contribute to Carmen's emotional well-being.

**Maintaining open communication and trust**

Another vital aspect of implementing AAL technology in Carmen's care plan is maintaining open communication. Carmen should be regularly informed about how the video monitoring system will work and what to expect. Her input should be valued, and she should feel involved in the decisions regarding her care. By having regular conversations with Carmen about her concerns and experiences with the technology, staff can build trust and ensure that she feels safe and respected.

Staff training is also essential to ensure that caregivers understand how to balance technology with personal care. Caregivers must be aware that while the video monitoring system improves safety, it cannot replace the emotional support and personal attention that residents like Carmen need. Training should emphasise that technology is a tool to assist caregivers, not a substitute for human interaction.

**Comprehensive fall prevention plan**

Given Carmen's history of falls, a comprehensive fall prevention plan should accompany the use of video monitoring technology. This plan could include safety modifications to her room, such as installing motion-activated lights, grab bars near the bed and in the bathroom, and anti-slip mats to reduce the risk of accidents.

Additionally, ensuring that Carmen has easy access to her mobility aids, such as a walker, can further enhance her safety and independence.

Physical therapy could also be introduced to strengthen Carmen's muscles and improve her balance, reducing the likelihood of future falls. By addressing the physical aspects of her care alongside the implementation of technology, the care home can take an integrated approach to improving Carmen's well-being.

**Conclusion**

Carmen's situation highlights the importance of balancing technology with personal care in a care home setting. While video monitoring technology can enhance Carmen's safety and prevent falls, it is essential to address her concerns about feeling isolated or abandoned. Open communication, regular personal interactions with caregivers, and active participation in the care home community are crucial to ensuring that Carmen feels safe, valued, and connected.

By combining technology with a comprehensive care plan that addresses both her physical and emotional needs, Carmen can continue to enjoy her time in the care home while feeling secure. The video monitoring system should be seen as a complement to the personal care she receives, not a replacement for the human connection that is so important to her well-being. With the right balance of safety measures, emotional support, and community engagement, Carmen can maintain her independence and quality of life in a supportive and caring environment.

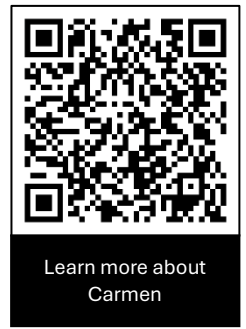

Learn more about Carmen

## 21. How do AAL technologies support daily living activities?

Activities of Daily Living (ADLs) are basic tasks like eating, bathing, dressing, using the toilet, and moving around. For older adults or individuals with disabilities, these tasks can become challenging due to physical or cognitive limitations. AAL technologies help overcome these challenges, enabling people to live independently and improve their quality of life.

AAL technologies support daily living by integrating smart devices and systems into the home. They assist with mobility through intelligent wheelchairs and walkers equipped with features like obstacle detection and automatic braking. These enhancements allow users to navigate their surroundings safely and confidently.

In personal care, AAL technologies aid with bathing and dressing. Voice-activated assistants provide reminders for grooming routines. They can also guide users through dressing steps, helping those with memory issues maintain personal hygiene.

Managing household tasks becomes easier with AAL technologies. Smart home systems automate lighting, heating, and appliance control, operable through voice commands or simple interfaces. This reduces physical strain and the risk of accidents, allowing users to manage their homes efficiently, even with limited mobility.

For nutrition and meal preparation, AAL technologies offer smart kitchen appliances that simplify cooking. Refrigerators can track inventory and remind users when food items are running low or nearing expiration. Meal planning apps suggest recipes based on dietary needs and available ingredients. Grocery delivery services coordinated through apps ensure access to fresh food without the need to visit stores.

Medication management is critical for many older adults and people with chronic conditions. AAL technologies provide electronic pill dispensers that alert users when it is time to take their medication and dispense the correct dose. They can also notify caregivers or family members if a dose is missed, helping prevent health complications due to missed medications.

Communication and social interaction are vital for mental and emotional well-being. AAL technologies facilitate this by offering devices with simplified interfaces, facilitating to stay in touch with family and friends. Video calling platforms enable face-to-face conversations, reducing feelings of isolation. Social networking apps designed for older adults help them connect with peers who have similar interests, fostering community engagement.

For cognitive support, AAL technologies offer digital calendars and reminder systems to help users keep track of appointments, events, and daily tasks. Voice-activated assistants provide prompts throughout the day, aiding in routine management and reducing confusion. These tools help maintain independence by supporting cognitive functions that might be declining.

Safety is a key concern in daily living. AAL technologies enhance safety through features like fall detection sensors. If a fall is detected, these sensors will automatically alert caregivers or emergency services. Environmental sensors monitor for hazards such as smoke, gas leaks, or unusual temperatures. They ensure that potential dangers are addressed promptly. Security systems with easy-to-use interfaces allow users to control door locks and monitor visitors, enhancing personal safety.

## 22. How do AAL technologies support mental health?

AAL technologies help people, particularly older adults and those with disabilities, live better by improving mental health and well-being. These technologies help by monitoring daily activities. They offer cognitive stimulation, enhance social connections, and respond to emergencies. By addressing emotional, social, and practical needs, AAL technologies promote independence and a higher quality of life.

AAL technologies play an important role in monitoring and early detection. Sensors and wearable devices can track daily activities. Sudden changes, like moving less or having trouble sleeping, might indicate depression, anxiety, or memory loss. By detecting these signs early, caregivers can intervene quickly, providing the necessary support before the issue worsens.

To keep the brain active and healthy, AAL technologies offer cognitive stimulation. These include games, puzzles, and memory exercises designed to sharpen the mind and help prevent cognitive decline, such as dementia or Alzheimer's disease. Regular mental activity is crucial for older adults to maintain their cognitive function.

Social interaction is another key feature of AAL technologies. Loneliness and isolation are significant risks for mental health, but AAL systems help users stay connected. Through video calls, online groups, and reminders to interact with loved ones, these technologies make it easier for people to maintain relationships and participate in social activities, even from home.

AAL technologies provide telehealth and remote counselling. They allow people to access medical and mental health care from home. Virtual meetings with doctors, therapists, or counsellors ensure that users receive necessary mental health support, particularly those who have difficulty travelling.

In cases of emergency, AAL systems can respond quickly. For instance, if someone is experiencing extreme anxiety or distress, the system can automatically call for help or notify a caregiver. This rapid response can prevent situations from escalating, offering peace of mind to users and their families.

AAL technologies also support caregivers' mental health by easing their responsibilities and providing peace of mind. The tools help with routine tasks, provide updates on the person's well-being, and allow caregivers to check in remotely. This reduces stress and gives caregivers a sense of control, helping them manage their duties more effectively. Knowing that their loved ones are safe and supported allows caregivers to focus on other aspects of their lives without constant worry.

Furthermore, personalised and environmental support is another benefit of AAL technologies. Users can receive reminders to take their medication or attend appointments. Smart home devices can create a calming atmosphere by adjusting lighting or playing soothing music, reducing anxiety, and promoting relaxation.

Additionally, AAL technologies offer access to mental health resources like mindfulness and relaxation exercises. By providing guided meditation, breathing techniques, and stress management tools, these technologies help users cope with anxiety and improve their emotional well-being. Educational content about mental health empowers individuals to understand and manage their feelings better.

## 23. What is the impact of AAL technologies on caregivers?

AAL technologies have a significant impact on caregivers by making their work easier, less stressful, and more efficient. These tools allow caregivers to remotely monitor the health and safety of older adults or people with disabilities. They can track key data like heart rate, blood pressure, and movement. With remote monitoring, caregivers can quickly respond if something goes wrong without needing to be physically present all the time. AAL technologies help caregivers by automating routine tasks, enhancing safety, providing flexibility, and improving communication with healthcare providers. This enables caregivers to deliver better care with less stress, helping them balance their caregiving duties with their personal lives.

One of the biggest advantages of AAL technologies is that they automate routine tasks. For example, they can remind individuals to take medication, attend appointments, or complete daily activities. Smart devices like pill dispensers ensure that the right medicine is taken at the right time, reducing the risk of mistakes. This automation allows caregivers to focus on more personal and complex care tasks, improving the overall quality of care. It also reduces the physical and mental burden on caregivers, preventing burnout and fatigue.

AAL technologies also enhance safety by preventing accidents such as falls. Devices that detect falls or unusual movements can send immediate alerts to caregivers or emergency services, ensuring a quick response. This gives caregivers peace of mind, knowing that they will be notified in case of an emergency. It reduces the need for constant supervision, enabling caregivers to step away while still feeling confident about the safety of the person they care for. Knowing that their loved ones are monitored and protected allows caregivers to relax and manage other aspects of their lives.

Remote monitoring tools also let caregivers check in from a distance, allowing them to manage other tasks and responsibilities more effectively. This flexibility helps prevent stress and burnout, which are common issues for caregivers. With the ability to stay connected even when they are not physically present, caregivers can balance their work and personal lives more easily. They can deal with their own needs, engage in social activities, or continue working, all while ensuring that the person they care for is safe and supported.

Additionally, AAL technologies improve communication between caregivers and healthcare professionals. The data collected by these systems can be shared with doctors and nurses, keeping them informed about the individual's health status. This collaborative approach ensures that the person receives better, more coordinated care, as medical professionals can adjust treatment plans based on real-time information. Caregivers can also receive guidance and support from healthcare professionals, enhancing their ability to provide effective care.

AAL technologies also support caregivers' mental health by easing their responsibilities and providing peace of mind. The tools help with routine tasks, provide updates on the person's well-being, and allow caregivers to check in remotely. This reduces stress and gives caregivers a sense of control, helping them manage their duties more effectively. Knowing that technology is assisting in monitoring and alerting them to any issues allows caregivers to rest easier, reducing anxiety and improving their well-being.

## 24. What information do AAL technologies collect about me?

AAL technologies collect a variety of personal and health-related information to provide personalised support and services. The specific data collected depends on the type of AAL technology being used, but generally includes:

- **Personal identification information:** Name, address, contact details, and date of birth to personalise the user's experience.
- **Health and medical data:** Vital signs (heart rate, blood pressure), medical history, medication schedules, and other health information to manage the individual's conditions.
- **Activity and behavioural data:** Information on daily routines and movements, like mobility patterns, exercise habits, or falls.
- **Location data:** Whereabouts tracked through GPS, especially useful for individuals with cognitive impairments.
- **Usage data:** How the user interacts with the technology—usage times, features accessed, and preferences—to enhance functionality.
- **Environmental data:** Details about the home environment (temperature, humidity, air quality) to improve safety and comfort.
- **Technical data:** Technical information like IP addresses and device identifiers to ensure compatibility and security.

Given the sensitive nature of this information, especially health and biometric data, AAL technologies are required to manage it securely and in compliance with data protection laws such as the General Data Protection Regulation (GDPR) in the European Union. These laws mandate that personal data must be processed lawfully, fairly, and transparently, and that appropriate security measures are in place to protect against unauthorised access or disclosure.

It is important to distinguish between three key types of data in AAL systems: **training data**, **input data**, and **output data**. **Training data** is used to teach AI systems how to function and make predictions. This data comes from a wide variety of sources and helps improve the accuracy of AI models. **Input data** refers to the information gathered by sensors or devices, such as your health stats. **Output data** is what the AI system produces, such as predictions about your health status or suggested treatments. Each type of data plays a crucial role in making the system work effectively.

As AI becomes more integrated into AAL technologies, laws like the **AI Act** are being developed to ensure that the data used for training these systems is accurate and free from bias. This is critical because biased or inaccurate data can lead to incorrect predictions or unfair treatment. The AI Act mandates strict standards for data quality and transparency to protect users' privacy and foster trust in AI-driven tools. Ensuring that health data is managed carefully, with measures like triple anonymisation, helps maintain confidentiality and protect users from potential misuse of their sensitive information.

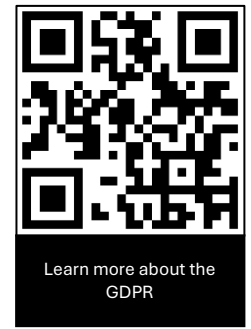

Learn more about the GDPR

## 25. How do I know what information is transmitted to my family and caregivers?

When using AAL technologies, it is important to understand what information is shared with your family or caregivers. These systems collect data through sensors, smart devices, or wearable gadgets to monitor your health, daily activities, and overall well-being. You usually have control over what data is shared, allowing you to customise the system based on your personal preferences and the needs of your caregivers.

The type of information shared depends on how the system is set up. Typically, safety-related data—like falls, emergencies, or unusual activity—is transmitted automatically. This helps your family or caregivers respond quickly if something goes wrong, ensuring you get help as soon as possible. In some cases, the system might track routine aspects of your life, such as medication schedules, movements around the house, or sleep patterns. While some people prefer to share only critical information, others might choose to share more detailed updates like health trends or daily activities.

One of the main advantages of AAL technologies is that they allow your loved ones to stay informed about your well-being without interrupting your daily life. This can provide peace of mind, knowing that your family or caregivers are aware of your safety without needing to constantly check-in. It helps you live more independently, with the reassurance that there is a safety net in place if needed.

However, it is important to be aware of the challenges, particularly around ethics and data privacy. Sharing personal, especially health-related, information raises concerns about how it is used and who has access to it. While most AAL systems offer privacy settings, there is always the risk that sensitive data could be accessed by unauthorised individuals or misused. This makes it crucial to understand the privacy policies of the technology you are using and ensure that your data is encrypted and well-protected against breaches.

There are also ethical considerations to keep in mind. Some people may feel uncomfortable with the idea of constant monitoring, even if it is for security reasons. This can lead to feelings of lost autonomy or being under surveillance. It is essential to give informed consent, knowing exactly what data is being collected and why. Your family and caregivers should respect your limits, ensuring that only the necessary information is shared and that you remain comfortable with the level of monitoring.

To stay informed about what information is transmitted, you can regularly review your AAL system's settings and privacy policies. Ask your provider for a clear explanation of what data is collected, how it is used, and who has access to it. Make sure you understand and agree with these terms before using the technology. Open communication with your family and caregivers is also important, so everyone is on the same page regarding your preferences and boundaries.

AAL technologies help keep you safe and connected with your family and caregivers. However, it is essential to manage your privacy settings carefully to ensure your data is protected and used responsibly while you benefit from the support these technologies provide.

**26. How do I know who has access to my data?**

When adopting AAL technologies, it is important to understand who has access to your data and how you can manage it. Privacy is a major concern with these technologies, as they often collect sensitive health and personal data. Knowing who has access to your data can help you maintain control and ensure that it is used responsibly.

Before using any AAL technology, start by reading the privacy policy and terms of use provided by the company. These documents explain how your data is collected, stored, and shared. They will tell you what kind of data is being gathered and with whom it may be shared, including third parties. You can find this information either on the company's website or within the app itself. By reading these documents carefully, you will understand what rights you have and how the company manages your data. If you have concerns, these documents help you decide if you are comfortable using the technology.

One of the most important aspects of managing your data privacy is user consent. In Europe, laws like the General Data Protection Regulation (GDPR) require companies to obtain your explicit consent before collecting or sharing your data. This means that when you start using a new device or app, you will often be asked to agree to the collection and use of your data through a pop-up window or consent form. Make sure to read these carefully and adjust your preferences to control who can access your information. Most systems will allow you to opt in or out of specific types of data collection.

Many AAL technologies offer settings that allow you to control who can view or access your data. These access controls give you the flexibility to decide whether to share your data with healthcare providers, caregivers, or others. For example, you may want to share health monitoring data with your doctor but not with a third-party service provider. By managing these settings, you can fine-tune how your data is shared and ensure that it is only accessible to the right people.

In addition to privacy settings, companies often offer further information about their data practices on their website. Some of them include a Frequently Asked Questions (FAQ) section that answers usual questions about data privacy and security. This can be a useful resource for understanding how the company protects your information and whether there are options for limiting data sharing.

If you still have concerns, do not hesitate to reach out to the company's customer support. Various AAL technology providers offer multiple ways to contact them, such as by phone, email, or even social media. Customer support can help clarify who has access to your data and assist you in making changes to your privacy settings.

Being proactive about understanding who has access to your data and how it is used is key to protecting your privacy. By reviewing privacy policies, managing consent, and adjusting access controls, you can use AAL technologies confidently while keeping your personal data safe.

## 27. What privacy risks are associated with AAL technologies?

AAL technologies provide many benefits by supporting the independence of older adults and people with disabilities, but they also pose significant privacy risks. These devices often gather detailed data on users' daily activities, such as sleep, movement, and even social interactions. This constant collection of data helps build a clear picture of a person's life, but it can feel intrusive. Many users may not fully understand the extent of the data being collected or how much of their personal life is being monitored. This raises serious concerns about their privacy and control over their own information.

One of the biggest privacy risks associated with AAL technologies is that they constantly collect data in the background. For example, smart sensors can track the user's location, activity levels, and interactions without the user actively engaging with the device. This creates a large volume of sensitive data, which, if not protected properly, can expose individuals to identity theft or data breaches. Weak security measures or outdated systems can also put this personal information at risk, leaving users vulnerable to cyber-attacks.

Another concern is how this data is processed and used. If the data collected by AAL devices is not managed properly, there is a risk that it could be used in ways that harm the user. For example, inaccurate data could lead to incorrect health advice or even faulty diagnoses, which could put a user's health at risk. There is also the risk of bias in how data is interpreted, which could result in unfair or inaccurate decisions being made about a person's care.

Data sharing is another potential privacy risk. Users may not always be aware of who has access to their data or how it is being shared. Companies providing AAL technologies may share data with third parties, such as healthcare providers, insurance companies, or even advertisers. This raises concerns about whether users are giving informed consent for their data to be shared and whether their data is being used in ways they are comfortable with.

To address these risks, companies developing AAL technologies must prioritise strong data protection measures. This includes using encryption and secure storage methods to protect sensitive information from unauthorised access. Developers must also ensure transparency by providing clear information to users about what data is being collected, how it will be used, and with whom it will be shared. Regulations like the General Data Protection Regulation (GDPR) help enforce these standards by requiring companies to obtain explicit consent from users before collecting or sharing their data.

In addition to data protection laws, there are growing efforts to improve the way AAL technologies use AI to process data. The AI Act, for example, aims to ensure that AI systems are trained on accurate, unbiased data and operate in a transparent and accountable way. This can help prevent biased outcomes and increase user trust in the technology. Explainable AI is another emerging field that focuses on making AI systems more understandable, so users and regulators can better grasp how decisions are made, which can further reduce privacy risks and improve trust in AAL systems.

## 28. What privacy settings are available in AAL technologies?

Privacy settings in AAL technologies are crucial for protecting personal data and ensuring users feel safe and secure while using the devices. These technologies often collect sensitive information, such as health data, location, or daily activities, so managing privacy settings is key. Here are some of the privacy settings commonly available in AAL technologies:

- **Data collection control:** AAL devices allow users to control what kind of data is collected. For example, some systems let you disable or enable certain features, like location tracking or health monitoring, depending on what you are comfortable sharing. You can often find these options within the device's privacy settings, allowing you to select which data points (like heart rate or step count) are shared with caregivers or healthcare providers.
- **Data sharing options:** AAL systems often provide data sharing options that let users control who can access their data. You can choose to share information with specific individuals, such as family members, caregivers, or healthcare professionals. For example, if you use a fall detection system, you can set it to alert only designated contacts in an emergency. Additionally, access permissions can be detailed, allowing you to control what each person can view. For instance, caregivers may have access to health data but not location information, ensuring greater privacy control.
- **Customisable alerts and notifications:** Some AAL technologies allow you to set custom alerts and notifications. This feature is particularly helpful in controlling what information is shared in real-time, such as sending fall alerts only under specific conditions. You can often choose the frequency of these alerts and who gets notified, keeping you in charge of your privacy.
- **Data retention settings:** AAL devices usually offer data retention settings, letting users control how long their data is stored. You can set time limits for how long your data remains in the system before it is automatically deleted. This helps protect your privacy by ensuring that old data does not stay in the system longer than necessary.
- **Anonymisation and de-identification:** Many AAL technologies offer anonymisation features, where your personal information is stripped from the data collected. This means that while the system can still gather useful insights (e.g., for research or system improvements), it cannot link the data to you. Anonymised data reduces the risk of privacy breaches, as your identity remains protected.
- **Multi-factor authentication:** Many AAL technologies include enhanced security measures like multi-factor authentication (MFA). This is an important security feature in many AAL technologies, ensuring that only authorised users can access sensitive data. MFA requires users to provide two or more verification methods before gaining access. Typically, this includes something the user knows (like a password), something they have (like a smartphone or token), or something they are (like fingerprint recognition). By using MFA, AAL technologies add an extra layer of protection, reducing the risk of unauthorised access, even if one authentication factor is compromised.

**29. What rights do I have regarding the usage of my data?**

Understanding your rights regarding your personal data is essential. In the EU, these rights are protected by the General Data Protection Regulation (GDPR). The GDPR gives you several important rights to protect your data:

1. **Right to Be Informed (Articles 12-14):** This gives you the right to know how your personal data is being collected, used, and shared. Companies must clearly inform you about the purpose of data collection and how long they will store your information. The privacy policy or terms of service from the company will typically provide this information.

2. **Right to Withdraw Consent (Article 7(3)):** If you have previously given your consent for a company to use your data, you can change your mind and withdraw that consent at any time. Companies must make it easy for you to withdraw your consent, and you should not face any penalties for doing so. Keep in mind that withdrawing consent does not affect the lawfulness of any processing that occurred before your consent was withdrawn.

3. **Right of Access (Article 15):** This gives you the right to know what data a company has collected about you. You can request access to your data, and companies are required to provide you with a copy of your personal information. They must also explain why they are processing your data and provide details about where it has been shared. You can usually expect to receive this information free of charge within one month of making the request.

4. **Right to Rectification (Article 16):** If the data a company holds about you is incorrect or incomplete, you have the right to request that it be corrected. This ensures that the information used about you is accurate and up-to-date, which is important for maintaining your privacy and preventing misuse of inaccurate data.

5. **Right to Erasure – "Right to be Forgotten" (Article 17):** In certain situations, you can request that your personal data be deleted. This might be applicable when the data is no longer needed for the purpose it was collected, or if you withdraw your consent. However, this right is not absolute—there may be legal reasons, such as public health concerns, that require companies to retain your data.

6. **Right to Restrict Processing (Article 18):** If you have concerns about the way your data is being used, you can request that its processing be restricted. This means that companies can store your data but cannot use it unless certain conditions are met. This is often used when the accuracy of the data is disputed or when legal issues are being resolved.

7. **Right to Object (Article 21):** You can object to how your data is being processed for certain purposes, such as direct marketing. Companies must stop processing your data unless they have strong legal reasons to continue.

8. **Right to Complain (Article 77):** If you think your data rights have been violated, you can file a complaint with a supervisory authority. This ensures that your concerns are properly addressed and investigated.

## 30. What should I know about data security in AAL technologies?

Data security in AAL technologies is important because these systems collect highly sensitive information, such as health metrics, daily routines, and personal data. This data needs to be protected from unauthorised access to ensure both privacy and user safety. Strong security measures, such as encryption, access control, and secure data storage, play a key role in maintaining the confidentiality, integrity, and availability of this information.

Encryption is essential in AAL systems. It ensures that data, whether at rest or in transit, is unreadable to anyone without the proper decryption key. This level of protection is especially important when transferring data from devices to central servers, such as when a wearable health monitor sends vital signs to a healthcare provider. Even if the data is intercepted, encryption makes it difficult for unauthorised individuals to access the information.

Access control is another critical element. It restricts who can view or edit personal data, using methods such as passwords, multi-factor authentication, or biometric data like fingerprints. Only authorised individuals, such as healthcare professionals, should have access to this sensitive information. This reduces the risk of misuse, particularly when handling data related to a person's health, which could be detrimental if used incorrectly.

Regular security audits are also vital for identifying vulnerabilities in AAL systems. These audits review security policies, test for potential weaknesses, and ensure that the latest updates are installed. With frequent changes in technology and evolving cyber threats, it is essential that security measures are continuously evaluated and strengthened.

Secure data storage is another fundamental component of data security. Sensitive information, including health records and personal details, should be stored in encrypted databases. This ensures that even if a breach occurs, the stored data remains protected and difficult for unauthorised parties to exploit.

In the event of a data breach, regulations such as the General Data Protection Regulation (GDPR) and the Cyber Resilience Act require companies to notify affected users and authorities promptly. Quick notification allows users to take steps to protect themselves. They can change passwords or monitor their accounts for suspicious activity. It also helps authorities to investigate the breach and implement measures to prevent similar incidents in the future.

Users of AAL technologies should be proactive in safeguarding their own data. This includes understanding the security measures implemented by the technology providers, asking questions about how data is stored and shared, and being vigilant about the potential risks of data breaches. By doing so, users can play a role in ensuring that their personal information is used responsibly and securely. Proper data security not only protects privacy. It also ensures that AAL technologies can be trusted to deliver the benefits they promise, such as improving health and safety for older adults and individuals with disabilities.

**Use Case Scenario 3: Carlos' experience as a caregiver using AAL technologies in a care home**

Carlos is a 47-year-old care home nurse in rural Spain. He has spent 16 years as a nurse, with the last seven years working in a care home where he has built strong relationships with both the residents and his colleagues. Carlos takes pride in his work and has created a positive, supportive atmosphere where everyone feels valued. However, recent changes at the care home have left him feeling uneasy.

The care home plans to install monitoring cameras in communal areas and residents' rooms for safety. While Carlos understands the purpose, he worries it may create a constant surveillance atmosphere, affecting trust between staff and residents. He is also concerned that residents might feel uncomfortable, possibly disrupting the sense of community he has worked hard to build.

**Carlos' concerns about cameras**

For Carlos, the priority is residents' safety and well-being while maintaining a strong sense of community. He has worked hard to create a comfortable environment where staff and residents interact smoothly. The introduction of cameras raises concerns that the atmosphere will shift to one of surveillance, causing tension. Carlos fears that constant monitoring could disrupt the friendly, relaxed space he has built, making both staff and residents feel uneasy under the pressure of being watched.

Carlos is worried about how the footage might be interpreted. He fears that management could misjudge situations, such as if a caregiver is busy and does not respond to a resident immediately. This might look like neglect in the video, even if the caregiver was occupied with another task. Misunderstandings like these could lead to unfair judgments, making Carlos' job more stressful.

**The benefits of monitoring systems**

Despite his concerns, he sees the benefits of installing monitoring systems. The cameras can help staff quickly respond to accidents like falls, preventing serious injuries. They also allow staff to monitor residents needing extra care without constant physical checks. This could make work more efficient, giving Carlos and his colleagues more time to focus on other important tasks.

Additionally, the cameras could reduce some of the stress that comes with constantly worrying about the safety of residents. With the cameras monitoring high-risk areas, staff might feel more confident that they will be alerted to any problems right away. This could give everyone some peace of mind, knowing that the system will help catch issues early on.

**Addressing Carlos' concerns**

To address Carlos' concerns, it is essential that the care home management creates an open line of communication. They should explain clearly how the cameras will work, what areas will be monitored, and how the footage will be used. Carlos and his colleagues need to understand that the cameras are there to enhance safety, not to monitor staff performance or create distrust.

Carlos needs reassurance that the data from the cameras will be used responsibly. Policies should ensure that footage is reviewed only for safety, not to criticise staff. He would feel more comfortable if access to the footage was limited to specific people, like the head nurse or emergency responders, in appropriate situations.

Training on how to use the system can ease some of Carlos' anxiety. If Carlos and the other staff members understand how the system works, how to access the footage in emergencies, and what to expect in case of incidents, they might feel more comfortable. Carlos should also be invited to give feedback during the process to make sure his concerns are considered.

**Maintaining a positive work environment**

Carlos is concerned that the cameras might harm the close, friendly atmosphere he has built. To address this, the care home should focus on maintaining a strong sense of community. Management could install cameras only in key areas like hallways and common spaces where accidents are more likely, rather than in private areas where staff and residents interact casually. This way, the balance between safety and maintaining a welcoming environment can be preserved.

Team-building activities and regular meetings where staff can share their concerns about the new system might help maintain a positive work environment. Carlos values teamwork and open communication, so giving him and his colleagues a chance to share their experiences with the cameras, voice concerns, and suggest improvements could be a clever way to keep everyone engaged and involved.

**Supporting Carlos' well-being**

Carlos' well-being is also at risk due to his anxiety about the cameras and potential changes at work, which could lead to burnout. To help staff like Carlos, the care home should consider offering support such as counselling or stress management services. This would help ease the transition and address concerns, ensuring that the staff can adapt to the new system without added stress.

Carlos enjoys spending time outdoors and engaging in physical activities during his free time. Management should encourage staff to take breaks, maintain a healthy work-life balance, and focus on their well-being outside of work to prevent burnout. This is especially important given the demanding nature of Carlos' job and the added stress of new technology in the workplace.

**Conclusion**

Carlos' concerns about the introduction of monitoring cameras in the care home are understandable. He values the positive, trusting atmosphere he has helped create and worries that the cameras will change this dynamic. However, he also understands the benefits these technologies can bring, especially in terms of resident safety.

By ensuring open communication, providing clear policies on how the cameras will be used, and involving Carlos and his colleagues in the decision-making process, the care home can address these concerns and make the transition to using AAL technologies smoother. With the right support, Carlos can continue to provide the high level of care he is known for, while benefiting from the safety improvements these technologies offer.

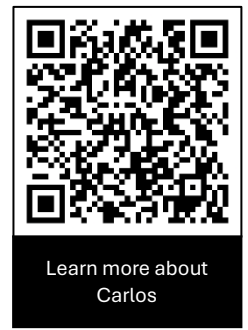

Learn more about Carlos

## 31. Where are my data stored?

AAL technologies collect large amounts of sensitive data, including medical information, behavioural patterns, and audio or video recordings. Where this data is stored depends on the system's design, with storage options including local devices, central servers, or remote data centres, often referred to as "the cloud." It is essential for users to be aware of where their data is stored and to take necessary precautions, such as understanding privacy policies and ensuring proper disposal of devices, to keep their information secure.

In simpler AAL devices, such as wearables, data is often stored locally on the device itself. Local storage limits the risks associated with transferring information over the internet and can offer greater control over data deletion. However, users must be cautious when disposing of these devices, ensuring that sensitive information is properly wiped to avoid potential exposure. For example, many devices store Wi-Fi credentials, and improper disposal could lead to unauthorised access to the user's home network.

More complex AAL systems often involve sharing data between devices to provide more integrated services. For example, a smart home system that monitors health may collect data from various sensors and send it to a central hub. This hub might process the data locally or send it to remote data centres for further analysis. Cloud storage is increasingly used in such cases because it allows for better scalability and remote access. The cloud enables large amounts of data to be stored and processed without the need for extensive local infrastructure.

While cloud storage offers many benefits, it also brings certain risks. Since the data is stored off-site, users must trust that the service provider has strong encryption and security measures in place. Data is often encrypted both during transmission and when stored in the cloud to protect it from unauthorised access. Encryption ensures that even if data is intercepted, it cannot be read without the appropriate decryption key.

Some AAL systems use a hybrid approach, combining both local and remote storage. For instance, data might initially be stored temporarily on a local device, where it can be quickly accessed for immediate use. Later, this data can be uploaded to the cloud for long-term storage or further analysis. This hybrid approach offers flexibility, balancing the need for quick, local access with the benefits of secure, centralised cloud storage. It also allows for data to be easily retrieved from the cloud when needed, without overloading local devices with large amounts of information.

Data in AAL technologies also exists in different forms, depending on how it is being used. Volatile memory stores data temporarily while the device is operating, but this data is lost when the device is turned off. Permanent memory stores data for longer periods, whether on the device itself or in remote data centres. As data moves through the system, it may exist in multiple places at once—in the device's memory, in communication channels, or in the cloud for long-term use.

**32. Who owns my data once it has been collected?**

When discussing data ownership in the context of AAL technologies, it is important to distinguish between personal and non-personal data. Personal data includes information that can identify you, like your name or health details, and you maintain rights over this data even after it is collected. These rights, such as access, correction, or deletion of your data, ensure that companies do not "own" your personal information in the traditional sense. You retain control over how it is used and shared, as outlined by privacy laws such as the GDPR.

Regarding non-personal data, such as anonymised information or statistics generated by devices, the European Data Act does not assign ownership to any party. Instead, it recognises that providers of smart devices, like AAL technologies, have exclusive control over the data generated by these devices. To address this, the Data Act introduces rules to ensure that data generated by devices and services can be shared more easily, while still protecting the interests of both data subjects and businesses. The act grants several rights to users:

1. **Right to Be Informed (Article 3):** Before using any AAL device or service, companies must clearly explain how your data will be managed. This includes the type of data collected, whether it is gathered continuously or at specific times, where it will be stored (locally or in the cloud), and how you can access it. This information should be simple and easy to understand, so you can be fully aware of how your data is handled.

2. **Right to Access and Use Data (Articles 3 and 4):** You have the right to access the data generated by your AAL devices. This allows you to view and use the data, such as health or activity information. You can share it with healthcare providers or use it for personal tracking. This right ensures that device manufacturers cannot limit your access to your own data.

3. **Right to Share Data (Article 5):** You have the right to share your data with third parties, like doctors or caregivers. This ensures you are not limited to using only one service provider. It encourages innovation by allowing you to switch services or combine data from different sources. Sharing data also helps personalise and integrate services across platforms.

4. **Right to Data Portability (Articles 4 and 5):** The Data Act, like GDPR, includes the right to data portability. This allows you to transfer your data from one service to another easily. For example, if you switch from one AAL system to another, you can move your health data without losing important information. This right helps ensure continuity of care and prevents any one provider from locking in your data.

5. **Right to Complain (Article 37):** If you believe your data rights have been violated, you can file a formal complaint with a supervisory authority. This ensures your concerns are investigated by an independent body, and action is taken to address the issue. Filing a complaint provides an important safeguard, reassuring users that any misuse or mishandling of their data will be taken seriously and that their rights are protected.

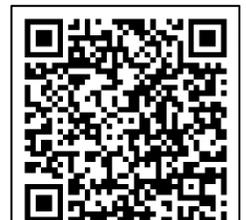

Learn more about the European Data Act

**33. Will my data be sold for commercial purposes?**

When using AAL technologies, it is important to understand how your data may be used, including whether it will be sold for commercial purposes. Companies collect user data for various reasons, such as improving services, personalising user experiences, and targeted marketing. To make informed decisions about your privacy, consider the following key points:

1. **Privacy policy and terms of use:** Every company providing AAL technologies must outline its data-handling practices in its privacy policy and terms of use. These documents should explain whether the company sells or shares your data for commercial purposes. It is important to review these documents carefully, which are usually available on the company's website or within the app. By reading them, you can understand how your data is used and whether it might be shared with third parties for marketing or commercial purposes.

2. **Regulatory environment:** In many countries, data protection laws such as the General Data Protection Regulation (GDPR) in Europe provide strong safeguards for users' data. Under the GDPR, companies are required to inform users if their data will be sold or shared for commercial purposes, and users have the right to opt-out. This regulation also mandates that companies cannot use personal data for purposes other than those originally specified when the data was collected, ensuring transparency in data usage. Regulations like GDPR hold companies accountable and give users control over their personal data.

3. **User consent:** In many cases, companies must obtain explicit consent from users before they can sell or share personal data. This consent is often given when users accept privacy policies or agree to cookie usage when first setting up a device or app. Consent banners or pop-ups allow users to agree to specific uses of their data, such as marketing or commercial sharing. Importantly, regulations like GDPR give users the right to withdraw their consent at any time, meaning you can change your mind and stop the company from selling or sharing your data.

4. **De-identified data:** Some companies may sell or share de-identified, or anonymised, data. This means the data is stripped of personal identifiers, making it difficult to trace back to an individual. While anonymised data is considered safer, it is still important to know how the company anonymises and uses this data. Privacy policies usually specify if anonymised data is being shared or sold.

5. **Managing your data:** To better control how your data is used, you can review privacy settings and consent options available within the app or device. Some AAL technologies allow users to limit the sharing of their data or opt out of certain practices, like targeted marketing. By adjusting these settings, you can protect your privacy while still using the technology.

Reviewing privacy policies, knowing the regulations in place, and making informed decisions about data sharing are all essential to ensuring your data remains secure and is not sold without your knowledge.

**34. What ethical issues arise with AAL technologies?**

There are, at least, five key ethical concerns about AAL technologies. They are autonomy, fidelity, beneficence, non-maleficence, and justice. These issues require careful consideration to ensure that AAL technologies benefit users without compromising their rights, well-being, or fairness.

1. **Autonomy:** While AAL technologies can enhance autonomy by supporting independent living, they can also risk reducing a user's control over their life. Some people may find AAL technologies intrusive due to constant monitoring or automation. For example, robot companions or pets might feel infantilizing to some older adults, undermining their dignity. It is important to involve users in decision-making processes to preserve their autonomy and prevent over-reliance on technology, which could cause them to lose essential skills. For example, if a smart heating system fails and the user cannot manually control it, this could create safety risks. Additionally, frequent automated alerts might desensitise users and caregivers, leading them to overlook important messages.

2. **Beneficence:** The principle of beneficence obligates healthcare providers and caregivers to use AAL technologies in a way that benefits the user's physical and mental well-being. Regular assessments of the user's preferences and experiences are necessary to ensure that AAL technologies truly support their recovery, autonomy, and dignity. These technologies should be used as tools to improve the user's quality of life until no further benefit can be achieved.

3. **Non-maleficence:** The principle of "do no harm" is critical when considering both the psychological and physical impacts of AAL technologies. These technologies enable remote care and monitoring. But they can increase loneliness if not balanced with human interaction. The reduction in in-person care may harm users' emotional well-being and social support, making it important to manage the integration of AAL solutions thoughtfully.

4. **Justice:** Justice in the use of AAL technologies involves fair treatment, respect for legal rights, and equitable access. AAL can provide great benefits. However, the costs and required digital skills may be too high for some, especially those from low socioeconomic backgrounds. This could widen the gap in healthcare access and create unequal opportunities for users across different communities, regions, and countries. Addressing these disparities is essential. It is vital to ensure that everyone, regardless of their wealth or technological expertise, benefits from AAL technologies.

5. **Fidelity and Security:** AAL technologies also raise concerns about security, such as the potential for hacking, surveillance, or data misuse. Sensors and cameras collect sensitive personal data, increasing the risk of unauthorised access. To address these issues, strict data protection measures and clear responsibilities between users, families, and system providers are essential. Fully informed consent is crucial to ensure that older adults understand how their data will be used. In addition, AAL technologies must not foster mistrust by making users feel targeted by stereotypes or unwanted advertisements. Users must trust both the technology and their relationship with the machine.

## 35. How do I establish which AAL technologies might be best for me?

If you are considering AAL technology, it is essential to evaluate a few important aspects to ensure you select the best option for your needs. Here is how you can approach the decision-making process:

1. **Identify your needs and objectives:** Start by assessing the specific tasks or activities where you need support. This might include health monitoring, mobility assistance, or help with daily routines. Consulting with a healthcare provider or caregiver can help clarify your needs and determine which AAL technologies can address them most effectively.
2. **Evaluate your home environment:** It is also important to make sure your home can accommodate the technology. Consider factors such as:
   - **Integration with Existing Technology**: Check if the AAL system can connect with other smart devices in your home, like lights, thermostats, or security systems.
   - **Reliable Internet Connection**: Many AAL devices rely on a steady internet connection for monitoring and alerts, so ensure your home has strong connectivity.
   - **Access to Electricity**: AAL systems often require constant power, so having easy access to outlets is essential.
   - **Space Availability**: Ensure there is enough room for devices, sensors, or furniture adjustments. Movable furniture or space expansion might be necessary to install the system properly.
3. **Budget:** Consider how much you are willing or able to spend. Look at the initial cost of the technology, as well as any ongoing expenses such as maintenance or subscription fees. Make sure your budget aligns with the type of AAL system you want to invest in.
4. **Assess Usability:** Determine if the technology will be easy for you to use. This includes:
   - **Ease of Use:** Check if the device is user-friendly and comes with clear instructions. Look for technology that can be operated without requiring too much technical knowledge or experience.
   - **Training Needs:** If the AAL system is complex, you might need some training to use it effectively. Consider whether you, or the person who will use the technology, will need help understanding and operating it.

Once you have reviewed your needs, home environment, budget, and usability, you will be better prepared to explore the AAL technologies that fit your situation. Being thorough in your evaluation ensures you choose the right system to enhance your quality of life and maintain your independence. The following checklist provides a guide to help you assess various technologies across different areas, from health monitoring to personal care. It helps identify the most suitable options for your needs while considering ease of use, costs, and integration with your home setup. Use it as a starting point to explore and compare your options.

**AAL Technology Checklist**

| Category | Example | Relevant technology | Selected (✓) |
|---|---|---|---|
| **Daily activities** | | | |
| Mobility issues | Difficulty moving around the home | • Automated or motorised furniture<br>• Intelligent wheelchairs<br>• Voice-controlled devices | |
| Health monitoring | Need to track vital signs or manage health conditions | • Wearable health monitors (e.g., heart rate monitors, glucose trackers)<br>• Remote health monitoring systems | |
| Cognitive support | Difficulty with memory or cognitive tasks | • Smart calendars and reminders<br>• Voice-activated assistants | |
| Safety | Risk of falls or need for home security | • Fall detection systems<br>• Smart locks and security cameras | |
| Household management | Difficulty with managing household chores | • Automated vacuum cleaners<br>• Smart dishwashers<br>• Voice-controlled home assistants<br>• Automatic pet feeders<br>• Smart thermostats | |
| Communication | Challenges with staying in touch with others | • Video communication tools (e.g., tablets with video calling)<br>• Social media access<br>• Communication boards or apps<br>• Hearing aids with Bluetooth | |
| Personal care | Difficulty with personal hygiene or cleaning | • Smart shower systems<br>• Automated bath lifts<br>• Electric toothbrushes with timers<br>• Adjustable-height sinks and mirrors | |
| Entertainment and leisure | Difficulty engaging in hobbies or leisure activities | • Voice-controlled entertainment systems<br>• Automated hobbies (e.g., robotic garden care)<br>• Digital streaming services<br>• Interactive gaming systems | |
| **Evaluate your living environment** | | | |
| Home layout | Large home with multiple rooms | • Automated lighting systems<br>• Smart home automation systems | |
| Existing infrastructure | Home with existing smart technology | • Integration with existing smart hubs or voice assistants<br>• Compatible devices that work with your current setup | |
| **Budget and cost** | | | |
| Affordability | Limited budget for technology | • Research affordable options<br>• Consider devices with lower upfront costs and minimal ongoing expenses | |
| **Usability** | | | |
| Ease of use | Need technology that is easy to operate | • User-friendly interfaces<br>• Devices with clear manuals and customer support | |
| Training needs | Complex technology that requires learning | • Training programs or tutorials<br>• Devices with hands-on training or easy-to-follow instructions | |

**36. What happens if my needs change?**

If your needs change, AAL technologies are designed to adapt and continue supporting you. One of the main advantages of AAL systems is their flexibility, allowing them to be adjusted or expanded as your health or living situation evolves. Whether you need more assistance with mobility, health monitoring, or daily tasks, these technologies can be updated to ensure you receive the right care and support. AAL systems are flexible and can be adjusted to meet your evolving needs, helping you maintain your independence and safety while receiving the care you require.

As your needs change, caregivers or healthcare professionals can modify the settings of your AAL system. For example, if you begin to experience mobility issues, additional tools such as smart walkers or voice-activated systems can be introduced. These devices help you move around more safely and independently. Other technologies, like automated lighting or motorised furniture, can also be added to make your home environment easier to navigate, reducing the risk of accidents.

If your health condition changes, AAL technologies can help manage these new challenges. For instance, if you develop a chronic condition like diabetes or heart disease, AAL devices can be updated to monitor vital signs such as blood sugar levels, heart rate, or blood pressure. These devices can send alerts to your caregivers or healthcare providers if something unusual happens, ensuring timely intervention. This helps keep you safe and allows your health to be monitored without the need for constant visits to a doctor.

In addition to health monitoring, AAL systems can provide more personalised reminders and support if you need help with daily tasks. If you start requiring more assistance with medication, appointments, or personal care routines, the system can be programmed to send you reminders at the right time. For example, smart pill dispensers can be set to remind you to take your medication and ensure the correct dosage is taken. If needed, these systems can also notify family members or caregivers to provide further support.

It is important to regularly assess your needs with caregivers or healthcare providers. Over time, your situation may evolve, requiring adjustments to your AAL system. By staying in touch with professionals, you ensure that your AAL technology is always tailored to your current needs. This helps maintain effective and reliable support. AAL technologies can be updated or expanded without the need for a full replacement, providing continuous, personalised care as your needs change. Regular reviews also offer opportunities to integrate new features or advancements, ensuring that you benefit from the latest innovations in AAL technology. This proactive approach helps maximise the usefulness of the system and ensures your safety, comfort, and independence over time.

Another benefit of AAL systems is that they help reduce the feeling of dependency. Even as your needs change, these technologies allow you to maintain your independence for as long as possible. By providing the right support when and where you need it, AAL systems help you continue living in your own home and managing your daily activities.

## 37. How do I obtain an AAL technology?

Obtaining AAL technology depends on your needs, budget, and location. AAL technologies include health monitors, fall detectors, smart home devices, and mobility aids. They aim to support older adults and people with disabilities in their daily lives. To find the right AAL technology, you should assess your needs, consult experts, research options, check for financial help, and decide whether to buy or lease. With the right approach, you can find technology that fits your lifestyle, enhances your independence, and improves your quality of life.

1. **Understand your specific needs:** The first step in getting AAL technology is knowing what kind of support you need. Do you require help with health monitoring, mobility, safety, or communication? For instance, a health monitoring system might be necessary if you have a chronic condition like diabetes. If safety is your concern, a fall detection device or smart home system might be better. Having a clear idea of what you need will help you choose the right technology.

2. **Consult with experts:** After identifying your needs, it is wise to consult a healthcare professional or caregiver. They can recommend the most suitable AAL technologies for your situation. Healthcare providers often have updated knowledge of the latest devices and can guide you in choosing options that improve your daily life. Some AAL technologies may also require a medical prescription or recommendation, especially those related to health monitoring. Consulting an expert ensures you select the right technology for your needs.

3. **Research available options:** There are many different AAL technologies available, so it is important to research different options. You can look online, visit local health equipment stores, or speak with companies that specialise in AAL technologies. There are many products to choose from. Make sure to compare the features, costs, and ease of use to find the best fit for you.

4. **Check for financial assistance or insurance coverage:** AAL technologies can be costly, but financial help might be available. Some health insurance plans or government programs cover part or all of certain AAL devices. It is worth checking with your insurance provider or local government for support. Charities or non-profits may also offer funding or discounts for people who need these devices but cannot afford them.

5. **Purchase or lease the device:** After choosing your AAL technology and checking for financial help, you can buy or lease it. Some devices are available for purchase, while more expensive or complex ones might be available for lease. Leasing is a good option if your needs may change or you want to try the technology before buying. This way, you can ensure that the device meets your requirements without making a long-term commitment.

6. **Installation and setup:** After obtaining the technology, you might need assistance with setup. Some devices are easy to install, while others, like health monitoring systems, may need professional help. Many companies offer installation services and will guide you on using the device properly. This ensures everything works as expected and provides the best support.

## 38. What training is required to use AAL technologies effectively?

To use AAL technologies effectively, it is important to have the right training to ensure users can get the most out of the devices and systems. The type and level of training required depend on both the complexity of the AAL technology and the user's familiarity with digital tools. In general, the training focuses on understanding how to use the technology, how to troubleshoot problems, and how to integrate these tools into everyday routines.

**Basic digital skills**

For many AAL technologies, users need a basic understanding of digital tools. This includes familiarity with operating devices like smartphones, tablets, and computers. Older adults or those with limited digital experience may require initial training to help them become comfortable with using these devices. Basic skills such as navigating apps, adjusting settings, and managing notifications are essential, as many AAL systems rely on smartphone apps or online platforms for controlling and monitoring devices. Training in this area can often be obtained through community centres, libraries, or online tutorials.

**Specific AAL technology training**

Once users have the basic digital skills, they will need specific training on the AAL technology they plan to use. This includes learning how to operate the device or system, understanding its functions, and knowing how to troubleshoot common problems. For example, if using a fall detection system, the user must know how to wear the device correctly and what to do when an alert is triggered. Many manufacturers provide manuals, instructional videos, or step-by-step guides to assist users. In some cases, trained technicians may visit the home to install and explain the system, ensuring it works properly and the user understands how to use it.

**Customisation and personalisation**

AAL technologies are often customisable, allowing users to set preferences that match their needs. For example, smart home systems might be set to adjust lighting and temperature based on the user's daily routine. Learning how to personalise these settings can significantly improve the user experience. This training typically involves learning how to adjust system preferences, set alarms and reminders, and integrate the AAL technology with other devices in the home. In many cases, online help desks or customer service teams can guide users through these processes.

**Ongoing support and learning**

A key part of the effective use of AAL technology is having access to ongoing support. Technology changes, and users' needs evolve over time, so ongoing training is often necessary. Many companies offer customer support hotlines, online resources, or even regular updates to ensure users continue to benefit from their devices. Family members and caregivers can also play a role by learning how to assist the user and keep the technology running smoothly.

Remember that education is also an opportunity to get to know new people and escape from routine. And very importantly, asking for help and explanations about technology to family and friends is another way of learning, and an excellent way to create topics in common to foster intergenerational solidarity.

### 39. How do I integrate multiple AAL technologies at home?

Integrating multiple AAL technologies at home involves connecting various devices to ensure they work seamlessly together. This process, known as "interoperability," is a key factor in creating a smart, supportive environment for users. Interoperability allows multiple devices from different manufacturers to communicate, share data, and perform tasks efficiently, enhancing the overall functionality of the AAL system.

Interoperability has been a long-standing challenge in technology, but efforts are being made to improve it. Both the tech industry and the open-source community are working hard to enable devices and systems to work together smoothly. Many platforms now allow users to manage and integrate various devices from different brands into one coherent system. For example, if you have a temperature sensor and a heating system connected to your smart home, you can set up automation that adjusts the temperature based on the time of day. During the day, you may want the temperature to be set at 22°C for comfort, while at night, it could automatically lower to 18°C for better sleep.

Automations are key to making AAL technologies work together. In many smart home platforms, an automation is essentially a rule that defines how devices should interact. For example, if the temperature falls below a certain level, the heating system will automatically turn on. This allows users to manage their environment without constantly adjusting settings manually. Additionally, these systems can send notifications to alert users if something unusual happens, such as a drop in temperature or if a door is left open. This feature ensures that users remain informed and can act promptly when necessary.

Many technology companies have developed their own ecosystems to promote interoperability, allowing their devices to work together seamlessly. These ecosystems are designed to ensure that the devices are compatible, making it easier for users to integrate various technologies within a single platform. By offering a unified approach, these ecosystems reduce the complexity of managing multiple devices from different manufacturers. Users can control everything from a central hub or app, ensuring that the devices offer similar features, respond to commands uniformly, and provide a smoother overall experience.

To achieve successful integration of AAL technologies at home, two main factors are crucial. First, devices must use a common data model. This means they should be able to understand and use shared information in a standard way. Second, devices must communicate using the same "language," known as a communication protocol, which allows them to exchange information and work together efficiently.

To integrate multiple AAL technologies in your home, ensure that your devices are compatible and can connect. A smart home platform or other ecosystems will help you link different devices. By ensuring your devices use a common data model and share a communication protocol, you can create automations that enhance your daily life, such as adjusting the heating or managing other home functions based on your preferences.

## 40. How do I choose the right AAL technology for my needs?

When choosing the right AAL technology, the process starts with identifying the primary user of the device and the environment where it will be used. Understanding who will use the technology, and in what setting, is essential for determining the effectiveness and appropriateness of the device.

The first step is to assess whether the primary user—whether it is you or a relative—is capable of learning and managing the technology. Cognitive abilities play a significant role in this decision. If the user is still lucid and cognitively healthy, they can learn to operate and manage the system independently. For example, they might manage settings, updates, or troubleshooting without assistance. However, if cognitive decline is a concern, family members or caregivers may need to take on this responsibility, so ease of use becomes crucial in selecting the right technology.

An important aspect of selecting AAL technology involves evaluating the individual's healthcare and social needs. This should be done with the help of healthcare professionals, such as general practitioners, gerontologists, physiotherapists, or nurses, who can assess physical well-being. In addition, social workers or psychologists can evaluate emotional and social needs. This comprehensive assessment helps identify the specific areas where AAL technologies can offer support, whether it is health monitoring, mobility assistance, or cognitive support tools. For instance, if a person has limited mobility, a device that assists with movement or safety might be a priority.

After identifying needs, explore available AAL technologies to find the best match. There are various resources, such as online catalogues, that offer detailed information on the wide range of AAL products available. It is beneficial to choose devices that are well-reviewed and have been evaluated by others in similar situations. Peer recommendations or products supported by trusted healthcare organisations can provide assurance that the technology will meet your needs.

The next consideration is where the device will be installed—whether in a home or an institutional setting. The environment plays a significant role in determining how effective AAL technology will be. In private homes, AAL technologies can improve safety and provide peace of mind. For example, monitoring systems can alert family members if a loved one with a neurodegenerative condition leaves the house or is at risk of falling. These devices enhance security while allowing the person to remain independent.

In institutions, such as care homes, different challenges may arise. Alarm technologies, which are useful in private homes, can sometimes overwhelm staff in a care facility, creating additional stress if not properly managed. In these cases, institutions may need to invest in additional staff to handle alarm responses. It is essential for care homes to weigh the costs and benefits of implementing these technologies, including independent evaluations of how they will affect both care and operational efficiency.

## Use Case Scenario 4: Isabel's challenge in implementing monitoring systems in a care home

Isabel is a care home manager in rural Spain with over 20 years of experience caring for older adults. She focuses on delivering high-quality care and creating a safe, supportive environment. Due to increasing emergency incidents, such as falls and wandering, Isabel is considering installing a monitoring camera system throughout the care home. This is particularly important for residents with dementia or mobility issues, as these incidents pose significant risks.

Isabel knows the system could enhance care by providing immediate emergency alerts and monitoring falls and resident movement. This would help prevent wandering, a concern in the care home. However, she also acknowledges the challenges, such as installation costs, maintenance, and legal obstacles that could complicate the process.

**Balancing technology and privacy concerns**

Isabel's top priority is the well-being and safety of her residents, but she is mindful of the privacy concerns that may arise. Residents, many of whom value their privacy and dignity, may feel uncomfortable being monitored constantly. Families may also express concerns about the ethical implications of video monitoring, worried that their loved ones could feel like they are under surveillance, which might affect their sense of freedom.

Isabel knows addressing concerns is key to implementing the system successfully. She plans to hold meetings with residents and their families to explain the system's purpose and use. Transparency is crucial, and Isabel believes open communication will ease concerns. She will stress that cameras will only monitor communal areas and high-risk zones, like hallways, and not invade private spaces such as bedrooms or bathrooms.

Isabel will establish guidelines on who can access the footage and how it will be handled to meet privacy regulations. Her priority is ensuring residents' safety without compromising privacy or dignity. Consent from residents and their families is crucial, and Isabel is committed to collaborating closely with them to find an ethical balance between safety and privacy.

**Gaining staff support and addressing resistance**

Isabel knows the staff are important to resident care and that monitoring cameras might cause unease, making them feel watched or judged. This could affect morale. Valuing teamwork and communication, Isabel is committed to addressing concerns head-on, ensuring staff understand the cameras' purpose and feel supported during the change.

To ease these worries, Isabel plans to involve the staff in the decision-making process. She will hold meetings where staff can express their concerns, ask questions, and offer input. By fostering an environment of transparency and collaboration, Isabel hopes to build trust and demonstrate that the monitoring system is not about scrutinizing staff but about ensuring residents' safety.

Isabel knows that proper training is essential to making the new system work. Staff must be taught how to use it, understand privacy rules, and feel confident that it benefits everyone. Isabel will organise training sessions to ensure the team is comfortable with the technology. She will also offer ongoing support and encourage feedback to help them adjust smoothly.

**Managing stress and preventing burnout**

As a care home manager, Isabel is under constant pressure to ensure that residents receive the best possible care while managing the operations of the facility. Implementing new technology, such as a monitoring camera system, adds an additional layer of responsibility. Isabel is mindful of the potential for stress and burnout, both for herself and her team.

To manage her workload effectively, Isabel has always maintained a strong sense of work-life balance. She values spending time with her family—her husband and two children—and engages in regular activities that help her recharge. Maintaining her personal well-being is essential for her to continue providing quality care to the residents.

Isabel knows her staff needs support as they adapt to modern technologies and increased responsibilities. She plans to hold regular check-ins to monitor stress levels and offer relaxation and team-building activities. By creating a supportive work environment, Isabel aims to keep her team motivated and engaged.

**Financial and legal considerations**

One of Isabel's biggest challenges in implementing the monitoring system is managing the costs. As a care home manager, she must carefully balance the budget. The cost of installing cameras, maintenance, and data management can strain resources. While she recognises these expenses are necessary for resident safety, she is also exploring funding options to reduce the financial burden. Isabel plans to reach out to local government agencies for grants or subsidies that support care homes implementing new safety technologies.

Isabel knows the legal and regulatory requirements, especially under the General Data Protection Regulation (GDPR), must be strictly followed. She will consult with legal experts to ensure compliance, including proper data encryption, limiting access to authorised personnel, and setting rules for how long footage is stored before secure deletion.

**Conclusion**

Isabel's decision to install a camera system in the care home is focused on improving resident safety. She is aware of the challenges, including privacy concerns, staff resistance, and financial and legal issues. To address these, she plans to prioritise both safety and dignity by maintaining open communication with residents, families, and staff, offering thorough training, and ensuring all legal requirements are met.

Ultimately, Isabel's approach reflects her dedication to providing high-quality care while embracing modern technology as a tool to enhance the quality of life for the residents. By addressing concerns thoughtfully and collaboratively, Isabel can ensure that the monitoring system is implemented in a way that benefits everyone involved, while maintaining the caring and supportive environment she has worked hard to create.

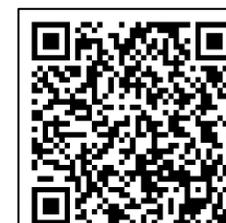
Learn more about Isabel

## 41. Is there a trusted organisation with which I can consult?

There are organisations that provide advice, support, and resources to help you choose the right AAL technologies that meet your needs. They assist in explaining how AAL systems work, offer information on costs, and help you find funding options. Consulting with these organisations ensures you receive reliable information and effective products to improve your quality of life. Non-profit organisations, government agencies, healthcare providers, and technology companies all offer valuable advice. This will ensure that you get the best possible support and guidance as you explore AAL technologies that suit your needs.

1. **Non-profit organisations:** Several non-profit organisations in Europe offer support for older adults and people with disabilities in finding appropriate AAL technologies. Organisations like Age Platform Europe or HelpAge International advocate for the well-being of older people, offering guidance on technology that can support independent living. These organisations often provide resources such as guides, workshops, and consultations to assist users in making informed decisions. Other European non-profits focus on improving the quality of life for people with disabilities and may also provide specific recommendations on AAL solutions tailored to individual needs.

2. **Government agencies:** Many European governments have national or regional bodies that provide information and assistance with AAL technologies. These agencies may be part of health or social services and can help guide you through the process of selecting and obtaining the right AAL devices. They also provide information on financial assistance or funding programs available to reduce the costs of these technologies. Reaching out to local health authorities or social service departments can help you find the necessary support.

3. **Healthcare providers:** Healthcare professionals are familiar with a wide range of AAL technologies and can recommend specific products that match your health and daily living needs. They can also ensure that the technologies you choose are safe and appropriate for your condition. Healthcare providers may be able to help with installation and offer training on how to use the technology effectively.

4. **Technology specialists and companies:** Some European companies specialise in AAL technologies. These companies often have customer service teams that can assist you with choosing the right devices. However, when consulting with a company, it is essential to verify its reputation. Check for reviews, certifications, and recommendations from healthcare organisations. They ensure the products and services are trustworthy.

5. **European AAL Programme:** The EU supports the AAL Programme. It aims to improve older adults' quality of life using AAL technologies. This programme funds research and development in this area and provides guidance on the latest advancements in AAL technologies. Their website has resources for those adopting AAL solutions. It also has insights into projects that could help older adults and caregivers.

## 42. Is there some kind of standard that I should be looking for?

AAL technologies are designed to improve the quality of life for older adults and individuals with disabilities, allowing them to live more independently. These technologies gather data to monitor health, activity levels, and environmental conditions. It is crucial to ensure the privacy and protection of this data. Several standards and frameworks exist to safeguard the use of AAL technologies and address concerns around data security and user privacy.

Key standards and regulations in this context include:

1. **ISO/IEC 24752 (Universal Remote Console - URC):** This standard focuses on interoperability and user interface adaptation. It ensures that different devices and systems can communicate effectively. This allows for seamless integration of AAL technologies in various settings.

2. **ISO/IEC 82304-1:** This standard is particularly relevant for AAL technologies as it ensures that the software used in these systems meets high standards for reliability and data protection. It provides assurance that software designed to assist older adults and individuals with disabilities is safe and dependable.

3. **IEC 62304:** This standard ensures that the software used in medical and AAL devices is safe and reliable, which is especially important given the vulnerable populations these technologies serve. It outlines the lifecycle requirements for medical device software, ensuring its safety and effectiveness.

4. **ISO/IEC 29100 (Privacy Framework):** This standard provides a framework for protecting privacy in Information and Communications Technology (ICT) systems. It is highly relevant to AAL technologies, as it addresses how personal information is managed and safeguarded, ensuring privacy is respected when collecting, processing, or sharing data.

Effective data protection in AAL technologies also involves compliance with specific frameworks designed to safeguard privacy:

1. **General Data Protection Regulation (GDPR):** In Europe, the GDPR mandates strict data protection measures. This includes data minimisation, ensuring that only necessary data is collected, and obtaining clear user consent before data is gathered or processed. GDPR also provides individuals with access rights to their data and requires notifications in case of data breaches. AAL technology providers must comply with GDPR to protect user data adequately and respect privacy.

2. **ISO/IEC 27001 (Information Security Management):** This standard helps AAL technology providers implement robust Information Security Management Systems (ISMS). ISO/IEC 27001 ensures the data collected by AAL technologies is confidential, intact, and available. It is a key standard for organisations that handle sensitive information.

By adhering to these standards and frameworks, AAL technology developers can ensure their systems are both secure and compliant with legal regulations. This builds trust and protects users' privacy while improving their quality of life.

## 43. Is there someone assisting me in using AAL technologies if something goes wrong?

When using AAL technologies, it is important to know how to get assistance if something goes wrong. Though these systems are designed to be user-friendly, issues can occasionally arise due to misconfigurations, software errors, or technical glitches. Knowing where to seek help can make the process less stressful and ensure that the technology continues to support your daily life.

The first point of contact when encountering a problem is usually the company's customer support or after-sales service team. They can help diagnose the issue and offer guidance on reconfiguring the device, updating the software, or scheduling a repair. Many AAL technology providers offer remote assistance, which allows technicians to troubleshoot the problem without needing to visit your home. This is particularly useful for older adults or individuals with mobility issues. Before purchasing any AAL system, ensure the company provides reliable and responsive after-sales support.

In some cases, AAL technologies are delivered through partnerships between public institutions and private companies. For example, your local government may collaborate with a technology provider to offer AAL services to residents. In this situation, local government workers or social services may assist in troubleshooting or act as intermediaries between you and the technology provider. These services can help coordinate repairs or replacements and may also offer advice on adapting the technology to your changing needs.

It is important to remember that AAL technologies are designed to put you in control. If you ever feel uncomfortable or think that the technology is not working as expected, you can turn off or unplug the device. Most systems are built with privacy features, allowing you to pause or stop data collection whenever you wish. Many devices come with an easy-to-access switch or button to override their operations, providing peace of mind that you remain in control of your environment.

In terms of safety, AAL devices are subject to the same regulations as any other household appliance. They are designed and evaluated to meet safety standards, ensuring they do not pose a risk in your home. However, if you notice any potential hazards, such as overheating or unusual noises, it is important to treat the device like any other electronic equipment. Switch it off immediately and contact the technology provider for repair or replacement. If the issue seems serious, like an electrical fire, you should contact emergency services to ensure your safety.

Additional support may be available for devices provided through public funding or government programs. Local authorities might offer assistance programs for repairs, updates, or replacements, ensuring you are never left without the help you need. This can be particularly useful in ensuring ongoing functionality and access to critical care services.

In conclusion, understanding where to get assistance and how to address issues with your AAL system is essential for maintaining the technology's effectiveness. Whether through the company's support services, local government assistance, or troubleshooting on your own, help is available to ensure these technologies continue to provide the support you need.

**44. Is it expensive to own AAL technologies?**

The cost of owning AAL technologies can vary widely depending on the specific type of devices and systems you choose. Some AAL technologies are relatively affordable, while others may come with more significant initial and ongoing costs. However, the potential long-term benefits, both in terms of independence and healthcare savings, often outweigh these expenses.

For basic AAL devices, such as fall detectors or medication reminders, the upfront cost is typically between €100 and €500. These types of devices generally have simple setups and do not require extensive maintenance. They are a cost-effective option for people looking to improve safety and daily living routines. However, more advanced AAL systems, such as health monitoring technologies that track vital signs or fully integrated home automation systems, can cost several thousand euros.

The initial cost of purchasing AAL technologies is only part of the total expense. Many AAL systems, particularly those that involve remote monitoring or cloud-based services, come with ongoing subscription fees. These fees cover the cost of data processing, cloud storage, and emergency response services. Subscriptions usually range from €10 to €50 per month, depending on the service level. For example, a subscription may include real-time health data monitoring, automatic alerts to caregivers, or regular well-being reports, which can add to the value of the technology.

Maintenance and replacement costs are also factors to consider. Devices like wearable sensors may require battery replacements or software updates to stay functional. In some cases, you may need to replace a device entirely after several years of use. These costs are generally manageable but should be factored into the overall financial planning for AAL technologies.

For those concerned about affordability, there are ways to offset some of these costs. In some European countries, government programs or healthcare providers offer financial assistance or subsidies for older adults or individuals with disabilities to help cover the cost of AAL devices. Non-profit organisations may also provide grants or funding options to make these technologies more accessible to those in need.

Renting or leasing AAL devices is another option, especially for people who may need the technology temporarily. This can be a more cost-effective alternative to purchasing, particularly for devices like health monitors that may only be needed for short periods. Renting also allows for flexibility, giving users the option to upgrade or change devices as their needs evolve without a large upfront investment.

In the long term, AAL technologies can help reduce healthcare costs by preventing accidents, reducing hospital admissions, and decreasing the need for in-person caregiving. For example, health monitoring devices can detect early warning signs of health issues. This allows for timely intervention and prevents more serious, costly medical treatments. By promoting independence and reducing reliance on caregivers, AAL technologies provide significant savings in both healthcare and long-term care settings.

### 45. Are there any after-sale costs that I might incur to continue using the AAL technology?

After purchasing AAL technologies, there are often additional costs to keep in mind to ensure their continued use. One of the most common ongoing expenses is maintaining an internet connection. Many AAL devices, especially those that connect to cloud services, require a stable broadband connection to function properly. If you do not have one, this will involve both installation fees and monthly payments to your Internet Service Provider (ISP). The cost depends on the provider and the type of service you choose, ranging from basic to high-speed connections, depending on the requirements of your AAL system.

Cloud service fees are another potential after-sale cost. Many AAL devices rely on cloud servers for data processing, storage, and remote monitoring. This means that, in addition to paying for the device, you may need to subscribe to a monthly or yearly plan that covers the cloud service's operational costs. These plans often include features such as data storage, real-time notifications, and remote access for caregivers. The exact fee depends on the type of device and services, so it is important to check these details before committing to a particular technology.

Maintenance is another factor to consider. Some AAL devices, like motion sensors or health monitoring tools, may require occasional upkeep, such as battery replacements. While batteries are inexpensive, this is still a recurring cost that you should account for. Similarly, for devices that run continuously, such as smart cameras, there will be a minor increase in your electricity consumption. Though minor, these costs can add up over time.

Software updates are often essential for AAL technologies to function efficiently and securely. Many providers include these updates as part of their ongoing service, but in some cases, there may be an additional fee to ensure the technology stays up-to-date. These updates help fix bugs, enhance functionality, and address security vulnerabilities, so they are crucial for keeping the system running smoothly.

Some AAL technologies also offer optional add-ons or premium features that can significantly enhance their functionality but may involve extra costs. These upgrades might include more advanced monitoring capabilities, such as detailed tracking of health metrics or additional safety features. Users might also choose to integrate their AAL systems with broader smart home networks, allowing for seamless control of various household devices. Other premium services could provide access to more detailed, real-time health reports or personalised support options, offering added convenience and tailored care for specific needs.

Lastly, it is worth exploring whether financial aid or subsidies are available to help cover these after-sale costs. Many local governments, social services, or non-profit organisations provide funding for older adults or individuals with disabilities to adopt and maintain AAL technologies. This assistance can help offset the costs of internet services, cloud subscriptions, and ongoing maintenance. It is advisable to consult with social services, healthcare providers, or welfare officers to find out what financial support might be available in your region.

## 46. How can I purchase or acquire AAL technologies through my health service?

Access to AAL technologies varies by country and healthcare system. In some countries, healthcare is funded by taxes and managed by authorities, while others use insurance systems. Whether AAL technologies are reimbursed depends on government policies at different levels. If these technologies are not covered, patients or residents will need to cover the costs themselves.

There are two primary areas where AAL technologies can be acquired: for home use and in residential care settings.

**AAL technologies for home use**: AAL technologies installed at home support independent living and enable older adults to age in place. Common examples include alarm and fall detection systems, which can be set up by healthcare or home care organisations. These systems often include alarm buttons or sensors that can call for assistance anytime. Often, the costs of these systems are covered by the healthcare organisation and reimbursed by the health authorities or insurance companies. Other technologies, like remote heart monitoring or blood pressure surveillance, are provided by healthcare organisations for use at home. Additionally, municipalities often supply equipment such as hoists and electric wheelchairs. This helps individuals maintain their independence. The patient usually pays a small personal contribution for these devices.

However, some AAL technologies are not covered by healthcare organisations. They must be purchased privately. These can include certain home support technologies available in specialised shops. Although the user bears these costs, they may be eligible for tax deductions as healthcare expenses, allowing individuals to reclaim part of the cost during tax filing.

**AAL Technologies in Residential Care Settings**: In nursing homes and sheltered housing, AAL technologies are typically provided by the healthcare organisation responsible for the care of residents. These technologies include alarm systems, fall and wandering detection systems, health monitoring devices, and tools for tracking sleep patterns. In many countries, healthcare organisations are allocated a budget for each patient. It covers personal care, medical treatments, physical therapy, and accommodation. The installation and use of AAL technologies are usually included within this budget.

In some cases, healthcare systems may offer additional financial support through special eHealth or AAL pilot projects. These pilots often target specific categories of patients who require extra treatment and monitoring, making them eligible for AAL technology support.

Whether AAL technologies are covered by healthcare systems depends on the policies and funding structures in each country. For those technologies not covered, individuals may have to fund them out of pocket. However, there may be opportunities for reimbursement through tax claims or special healthcare schemes. Individuals and their caregivers should contact local authorities, healthcare providers, or insurance companies to determine what financial assistance is available for AAL technologies.

**47. How can AAL technologies reduce healthcare costs?**

Healthcare costs mainly come from personnel expenses, which account for around 80% of the total, though this varies by sector. Hospitals tend to spend more on equipment and medications, while long-term care and elderly care depend more on staff. AAL technologies have the potential to reduce these costs but also bring additional expenses, such as software updates, maintenance, and IT support. It is also important to consider the environmental impact of producing these technologies, particularly with the use of rare materials.

To effectively reduce healthcare costs, AAL technologies should aim to decrease dependence on personnel. They should also consider the costs of implementing and maintaining these systems. Achieving these outcomes can make AAL technologies a financially viable option. Here are some ways in which AAL technologies can help reduce healthcare costs:

- **Promoting user independence:** AAL technologies can reduce the need for staff by enabling older adults and patients to manage daily activities on their own. For instance, automated medication dispensers and reminders allow users to handle their medication without staff intervention. Other examples include voice-controlled home facilities, robot vacuums, and smart appliances. They enable individuals to maintain their routines autonomously.

- **Remote monitoring:** Wearable devices and health sensors allow caregivers and healthcare providers to track vital signs and daily activities remotely. This decreases the need for in-person visits. This also helps detect potential health issues early, preventing costly emergency interventions.

- **Mobility aids:** Technologies such as sensor-equipped walkers and wheelchairs can help users move around safely and independently. Wearable devices that provide outdoor guidance further support mobility. Shopping delivery apps can also reduce the need for staff to assist with errands, allowing users to order goods themselves.

- **Safety features:** Safety is a major concern for older adults. Technologies like fall detection systems, automated lighting, and emergency call devices give users peace of mind. This reduces the need for staff intervention. These systems can alert caregivers or family members in case of emergencies, ensuring timely responses without the need for continuous monitoring.

- **Personal care:** Robotics can reduce the need for staff to provide personal care, such as assisting with bathing or transferring individuals from bed to chair. Automated aids, like shower systems and adjustable furniture, help users care for themselves. They reduce the need for staff.

- **Reducing social isolation:** Social isolation can lead to health problems, and increasing healthcare demands. AAL technologies can help reduce isolation by offering communication platforms that allow users to stay connected with family, friends, or support groups. By preventing loneliness-related health issues, these technologies reduce the need for additional healthcare interventions.

## 48. What funding options are available for AAL technologies?

Funding options for AAL technologies vary depending on the country, region, and specific healthcare system. However, there are several common sources that individuals and families can explore to help cover the cost of AAL technologies. These options include government programs, private insurance, charitable organisations, and sometimes even specific grants or subsidies.

1. **Government funding and subsidies:** Many countries provide funding for AAL technologies through healthcare or social care programs. This support may include subsidies for home care equipment like fall detectors, medication dispensers, and mobility aids. Some governments offer programs to help older adults live independently for longer. These subsidies, available through healthcare or welfare systems, can help with the cost of purchasing and maintaining AAL devices.

   In Europe, many public health systems fund AAL technologies to reduce hospitalisations, support home care, and enhance older adults' quality of life. Some regions also offer local grants for home adaptations and smart devices.

2. **Private insurance coverage:** Private health insurance may sometimes cover AAL technologies, offering partial or full reimbursement for assistive devices and remote monitoring systems. It is important to review your insurance policy to check if AAL devices are included. Coverage often depends on whether the technology is considered medically necessary, and some insurers may require a doctor's recommendation.

3. **Grants and charitable organisations:** Non-profit organisations, charities, and foundations often help fund AAL technologies for older adults or people with disabilities. These grants typically focus on individuals with limited financial resources and may cover both the initial device costs and ongoing maintenance. Such organisations aim to provide tools that promote independent living. Some charities provide financial help for smart home technologies, mobility aids, or monitoring systems to assist caregivers. This support is valuable for families who may not qualify for government assistance but still need help affording these technologies.

4. **Public-private partnerships:** In some regions, public-private partnerships help fund AAL technologies. These collaborations between governments and private companies aim to make assistive devices more affordable for older adults. For instance, a government might provide funding or tax incentives to a company developing smart home technologies for elderly care, making these solutions more accessible.

5. **Tax deductions and benefits:** In some countries, costs related to healthcare or home adaptations, including AAL technologies, may be tax-deductible. This means that individuals can claim expenses for purchasing and maintaining these devices when filing their taxes, reducing the financial burden.

**49. How do I stay informed about new AAL technologies?**

AAL technologies can enhance independence, safety, and quality of life, but keeping up with advancements can be challenging. Here are several ways older people and their caregivers can stay updated on the latest developments:

1. **Subscribing to newsletters and websites:** Many organisations that focus on elder care or assistive living provide free newsletters. These newsletters often include updates on the latest AAL technologies, research developments, and product releases. Subscribing to reputable sources such as healthcare organisations, universities, and technology providers is an easy way to receive regular information. Some websites from senior care organisations often offer tips, guides, and product reviews tailored to the needs of older adults and their caregivers.

2. **Attending community workshops and seminars:** Local community centres, libraries, and senior organisations often hold workshops and seminars to introduce older people and caregivers to new technologies. These events can provide hands-on experience with new AAL devices, as well as guidance on how to use them. In addition, webinars are increasingly popular and accessible from home. Many healthcare organisations and tech companies host free webinars that discuss the benefits and usage of the latest AAL innovations. These events can also provide an opportunity to ask questions and get expert advice.

3. **Joining support groups and online communities:** Support groups, both online and offline, are valuable resources for staying informed about AAL technologies. Many older adults and caregivers share their experiences and recommendations regarding new solutions in these groups. Specialised forums for caregivers offer spaces where people can discuss their challenges, share knowledge, and discover new AAL devices. Engaging with others in these communities can also help caregivers find practical insights from real-life experiences.

4. **Collaborating with healthcare providers:** Healthcare professionals often recommend AAL technologies to improve patient care. Regular check-ups and consultations provide a good opportunity to ask about the latest devices that may support health and safety. Medical professionals stay informed about the newest technologies through continuing education, making them a reliable source of information.

5. **Visiting technology and senior expos:** Attending expos focused on healthcare technology or senior living is another great way to stay informed. These expos showcase the latest innovations in AAL, often with live demonstrations. While attending in-person may be challenging for some, virtual expos have become increasingly common, providing an opportunity to explore new products from home.

6. **Participating in training courses:** Some organisations offer training courses specifically designed for older adults and their caregivers to learn about using AAL technologies. These courses can cover basic digital literacy, how to use smart home devices, or how to manage health apps. Government agencies and non-profit organisations sometimes offer free or low-cost training to help older adults and caregivers feel confident using new technologies.

## 50. What future developments are expected in AAL technologies?

AAL technologies are rapidly evolving to better serve older adults and people with disabilities. As technology advances, AAL systems will become more effective, user-friendly, and accessible. Improvements in AI, smart homes, and communication tools will enhance health monitoring, safety, and independence, allowing people to live comfortably in their homes for longer. These advancements will enable smarter, more intuitive systems, leading to better integration and overall quality of life for users.

1. **More advanced artificial intelligence (AI):** One area of development is the use of artificial intelligence (AI) in AAL technologies. AI can make systems smarter by learning from users' habits and behaviours, predicting when help might be needed. It can detect early signs of health decline, enabling quicker interventions. AI can also improve personal assistants, making them more conversational and capable of handling complex tasks like managing health schedules and providing reminders, thus enhancing overall care.

2. **Improved health monitoring and diagnostics:** Future AAL technologies will offer even better health monitoring tools. Wearable devices, such as smartwatches or fitness trackers, will become more advanced, able to monitor a wider range of health indicators. These devices could check for heart disease, diabetes, and mental health issues. They would provide real-time feedback to users and healthcare providers. Also, AAL technologies may use advanced diagnostic tools to track subtle changes in a person's health. These tools can alert caregivers or doctors when intervention is needed.

3. **Enhanced integration with smart homes:** The future of AAL technologies will also involve greater integration with smart home systems. This means that more devices in a person's home will work together to create a fully automated environment. For example, AAL technologies could connect with smart appliances, lighting, and heating systems to create a comfortable and safe living space. If someone forgets to turn off the stove, the system could detect this and turn it off automatically. The integration of smart home features will help people live more independently and safely.

4. **Better communication tools:** Future developments in AAL technologies will make it easier for people to stay connected with family, friends, and caregivers. Improved video call systems and social networking platforms designed for older adults will help reduce feelings of loneliness and isolation. These tools will become simpler to use, with voice-activated assistants making it easier to make video calls or send messages. This will be especially relevant for people who may struggle with traditional technology, such as smartphones or computers.

5. **Increased affordability and accessibility:** AAL technologies are evolving. They should soon be cheaper and easier to access. Advances in manufacturing and technology will likely lower the costs of these devices, making them available to a wider audience. This will ensure that more people can benefit from AAL technologies, regardless of their financial situation.

**Acknowledgement**

The personas Carmen, Carlos, and Isabel, on which some of the use cases in this booklet are based, were developed by Tamara Mujirishvili. More information can be found at: Mujirishvili, T., & Florez-Revuelta, F. (2023). Understanding User Needs, Persona Scenarios for Privacy-Preserving Visual System Development. In *Assistive Technology: Shaping a Sustainable and Inclusive World* (pp. 97-104). IOS Press.